\documentclass[english,prd,superscriptaddress,nofootinbib,preprintnumbers]{revtex4} 
\usepackage{graphicx}% Include figure files
\usepackage{dcolumn}% Align table columns on decimal point
\usepackage{bm}% bold math
\usepackage{latexsym}
\usepackage{amsfonts}
\usepackage{amssymb}
\usepackage{amsmath}
\usepackage{bm}
\usepackage{color}

\def\be{\begin{equation}}
\def\ee{\end{equation}}
\def\ba{\begin{eqnarray}}
\def\ea{\end{eqnarray}}

\begin{document}

\preprint{KUNS-2459}

\title{Observational signatures of anisotropic inflationary models}

\author{Junko Ohashi}

\affiliation{Department of Physics, Faculty of Science, Tokyo University of Science, 
1-3, Kagurazaka, Shinjuku, Tokyo 162-8601, Japan}

\author{Jiro Soda}

\affiliation{Department of Physics, Kyoto University, Kyoto 606-8502, Japan}

\author{Shinji Tsujikawa}

\affiliation{Department of Physics, Faculty of Science, Tokyo University of Science, 1-3, Kagurazaka, Shinjuku, Tokyo 162-8601, Japan}

\date{\today}% It is always \today, today,

%===============================================================%
%************************* ABSTRACT ****************************%
%===============================================================%

\begin{abstract}

We study observational signatures of two classes of anisotropic 
inflationary models in which an inflaton field couples to 
(i) a vector kinetic term $F_{\mu \nu}F^{\mu \nu}$ and 
(ii) a two-form kinetic term $H_{\mu \nu \lambda}H^{\mu \nu \lambda}$.
We compute the corrections from the anisotropic sources to the 
power spectrum of gravitational waves as well as the two-point 
cross correlation between scalar and tensor perturbations. 
The signs of the anisotropic parameter $g_*$ are different 
depending on the vector and the two-form models, but 
the statistical anisotropies generally lead to 
a suppressed tensor-to-scalar ratio $r$ and a smaller scalar 
spectral index $n_s$ in both models. 
In the light of the recent  Planck bounds of $n_s$ and $r$,
we place observational constraints on several different inflaton 
potentials such as those in chaotic and natural inflation 
in the presence of anisotropic interactions.
In the two-form model we also find that there is no cross correlation 
between scalar and tensor perturbations, while in the vector 
model the cross correlation does not vanish.
The non-linear estimator $f_{\rm NL}$ of 
scalar non-Gaussianities in the two-form model
is generally smaller than that in the vector model for the same 
orders of $|g_*|$, so that the former is easier to be 
compatible with observational bounds of non-Gaussianities 
than the latter.

\end{abstract}


\maketitle

%===============================================================%
%************************ SECTION I ****************************%
%===============================================================%

%%%%%%%%%%%
\section{Introduction}
%%%%%%%%%%%

The measurements of Cosmic Microwave Background (CMB) temperature anisotropies 
and large-scale structures have significantly improved in accuracy over the last 
decade \cite{WMAP1,WMAP9,Rei,Atacama,LSS1,LSS2}. 
In particular, the recently released Planck data \cite{Adecosmo,Adeinf,Adenon} 
showed that the primordial power spectrum of curvature perturbations is 
slightly red-tilted from the exact scale-invariance.
This is consistent with the theoretical prediction of standard slow-roll inflation driven by 
a nearly flat potential of a scalar field 
$\phi$ (called ``inflaton'') \cite{oldper}.

While the WMAP and Planck data support the inflationary scenario overall, 
there are some anomalies in the data \cite{WMAP9,Adecosmo} 
which are difficult to be addressed in the context of single-field slow-roll inflation. 
One of them is broken rotational invariance of the CMB 
perturbations \cite{aniobser1,aniobser2,aniobser3}.
The power spectrum of curvature perturbations $\zeta$ 
with broken statistical isotropy can be expressed 
in the form \cite{Carroll}
\be
{\cal P}_{\zeta} ({\bm k})=
{\cal P}_{\zeta}^{(0)} (k)\,\left(1+g_*\cos^2 
\theta_{{\bm k},{\bm V}} \right)\,,
\label{anispe}
\ee
where ${\bm k}$ is the comoving wave number, ${\cal P}_{\zeta}^{(0)} (k)$
is the isotropic power spectrum, $g_*$ characterizes the deviation from 
the isotropy, ${\bm V}$ is a privileged direction close to the ecliptic poles, 
and $\theta_{{\bm k},{\bm V}}$ is the angle between ${\bm k}$ and ${\bm V}$.
{}From the WMAP data, Groeneboom {\it et al.} \cite{aniobser3} obtained the bound 
$g_*=0.29 \pm 0.031$ with the exclusion of $g_*=0$ at $9\sigma$ by 
including the CMB multipoles up to $\ell=400$. 
There is still a possibility that some systematic effect such as
the asymmetry of the instrument beam accounts for broken 
rotational invariance \cite{Hanson} \footnote{Two months after the 
initial submission of our paper, Kim and Komatsu \cite{Kim} obtained
the bound $g_*=0.002 \pm 0.016$ (68\%\,CL) by using the Planck data. 
This limit was derived after eliminating the asymmetry of the Planck beam  
and the Galactic foreground emission. While the isotropic power spectrum 
is consistent with the Planck data, the anisotropy of the order of $|g_*|=0.01$ is not 
excluded yet. In this paper we allow for the possibility that $|g_*|$ 
can be of the order of 0.1 without using the new bound of Kim and Komatsu.}.

If broken statistical isotropy is really present for primordial perturbations, 
we need to go beyond the slow-roll single-field inflationary scenario
to explain the origin of statistical anisotropies \cite{review1,review2}.
For the models in which the inflaton $\phi$ has a coupling with 
a vector kinetic term $F_{\mu \nu}F^{\mu \nu}$, there exists an 
attractor-like solution along which an anisotropic vector hair 
survives even during inflation \cite{Watanabe} 
(see Refs.~\cite{early} for early related works and Refs.~\cite{powerlaw,anipapers} 
for rich phenomenologies of anisotropic inflation). 
Even if the background energy density of the vector field is suppressed relative 
to that of the inflaton, the isotropic power spectrum ${\cal P}_{\zeta}^{(0)} (k)$ 
is modified to have the form (\ref{anispe}) with a negative 
anisotropic parameter $g_*$ \cite{Gum,Watanabe:2010fh,Bartolo}. 
Moreover the non-linear estimator $f_{\rm NL}$ of scalar non-Gaussianities 
can be as large as the order of 10 for the squeezed shape averaged over 
all directions of a wave number 
${\bm k}_3$ with ${\bm k}_1 \simeq -{\bm k}_2$ \cite{Bartolo,Shiraishi}
(see also Refs.~\cite{Barnaby} for the large non-Gaussianities generated 
by vector fields).

Recently, the present authors showed that anisotropic inflation can be 
also realized for the model in which the inflaton couples to a two-form field 
with the kinetic term $H_{\mu \nu \lambda}H^{\mu \nu \lambda}$ \cite{Ohashi} (see 
Ref.~\cite{Watanabe:2010fh} for an early proposal).
In this case, the anisotropic power spectrum is given by Eq.~(\ref{anispe})
with $g_*>0$. Since the sign of $g_*$ is opposite to that in the vector model, 
we can observationally distinguish between the two anisotropic 
inflationary scenarios.
In Ref.~\cite{Ohashi} the bispectrum and trispectrum 
of curvature perturbations have been also evaluated 
in the two-form model by using the interacting Hamiltonian 
picture \cite{Maldacena:2002vr,Bartolo}. 
It was shown that, in the strict squeezed limit, the non-linear estimator
$f_{\rm NL}$ vanishes and that, in the equilateral and enfolded limits, 
$f_{\rm NL}$ can be larger than the order of 10. 

In this paper we place observational constraints on two anisotropic inflation models 
based on the vector and the two-form fields in the light of the recent 
Planck data \cite{Adecosmo,Adeinf,Adenon}. 
We first derive the anisotropic power spectra of gravitational waves 
to evaluate the tensor-to-scalar ratio $r$ correctly.
Using the observational bounds of $r$ as well as the scalar spectral index $n_s$ 
constrained by the joint data analysis of Planck and other
measurements \cite{Adeinf,Kuro}, we test for several representative 
models such as chaotic and natural inflation 
in the presence of anisotropic corrections to the scalar and tensor 
power spectra.

We also compute the cross correlations between curvature perturbations and 
gravitational waves. While the cross correlation survives in the vector model, 
it vanishes in the two-form model. This property is useful to distinguish 
between the two anisotropic inflationary scenarios from the correlation between the observed 
temperature perturbation and the B-mode 
polarization (TB correlation) \cite{Watanabe2}.

We also revisit the estimation of the anisotropic scalar non-Gaussianities 
for several different shapes of momentum 
dependence (local, equilateral, and enfolded shapes) 
in both the vector and the two-form models.
In fact, we show that the local non-linear estimators in the strict 
squeezed limit (${\bm k}_3 \to 0$ with 
$\theta_{{\bm k}_1, {\bm k}_3} \to \pi/2$, $\theta_{{\bm k}_2, {\bm k}_3} \to \pi/2$) 
vanish in both models.
However, if we average over all directions of a wave number ${\bm k}_3$
with ${\bm k}_1 \simeq -{\bm k}_2$ for nearly squeezed shapes \cite{Bartolo}, 
we have non-zero values of $f_{\rm NL}$ for $|g_*|>0$. 
Taking this prescription, we show that the local non-linear estimator 
$f_{\rm NL}^{\rm local}$ in the two-form model is smaller than 
that in the vector model by one order of magnitude for 
the same order of $|g_*|$.

This paper is organized as follows.
In Sec.~\ref{backsec} we review the background dynamics of anisotropic inflation 
for both the vector and the two-form models.
In Sec.~\ref{perksec} we derive anisotropic corrections to the scalar/tensor power 
spectra and also evaluate the cross correlation between curvature perturbations and 
gravitational waves.
In Sec.~\ref{obssec} we put constraints on several different inflaton 
potentials by using the 68\,\%\,CL and 95\,\%\,CL observational contours
in the $(n_s, r)$ plane. 
In Sec.~\ref{nonsec} we compare two anisotropic inflationary models from 
the non-linear estimator $f_{\rm NL}$ of scalar non-Gaussianities.
Sec.~\ref{consec} is devoted to conclusions.

%===============================================================%
%************************ SECTION II ****************************%
%===============================================================%

%%%%%%%%%%%%%%%%%%
\section{inflation with anisotropy}
\label{backsec}
%%%%%%%%%%%%%%%%%%

In this section we briefly review the background dynamics of anisotropic inflation 
for two classes of models in which the inflaton field $\phi$ couples to 
(i) a vector field $A_{\mu}$ \cite{Watanabe} and 
(ii) a two-form field $B_{\mu \nu}$ \cite{Ohashi}.
For suitable choices of couplings, the energy densities of 
the fields $A_{\mu}$ and $B_{\mu \nu}$ can survive even 
during inflation. For details, we refer the reader 
to the review articles~\cite{review1,review2}.

%%%%%
\subsection{$f(\phi)^2F_{\mu \nu}F^{\mu \nu}$ model}
\label{vecsec}
%%%%%

Let us first discuss the vector model given by the action 
\begin{eqnarray}
S&=&\int d^4x\sqrt{-g}\left[ \frac{M_{\rm pl}^2}{2}R
-\frac{1}{2} \partial_\mu\phi \partial^{\mu}\phi
-V(\phi)-\frac{1}{4} f(\phi )^2 F_{\mu\nu}F^{\mu\nu}  
\right]\,,
\label{eq:action}
\end{eqnarray}
where $g$ is the determinant of the metric $g_{\mu \nu}$, 
$M_{\rm pl}$ is the reduced Planck mass, and 
$R$ is the scalar curvature.
$V(\phi)$ and $f(\phi)$ are a potential and a kinetic function 
of the inflaton $\phi$, respectively.
The field strength of the vector field is characterized by
\begin{equation}
F_{\mu\nu} = \partial_\mu A_\nu - \partial_\nu A_\mu \ .
\end{equation}

Now, we consider cosmological solutions in this system. 
Without loosing the generality, one can take $x$-axis 
for the direction of the vector field. 
Using the gauge invariance, we can express 
the vector field as 
\begin{equation}
A_{\mu}dx^{\mu} = v(t) dx \,,
\end{equation} 
where $v(t)$ is a function of the cosmic time $t$.
Even if initial inhomogeneities and isotropies in $v$ are 
present, it was shown that only the background field $v(t)$ 
survives during anisotropic inflation in the Bianchi type I 
Universe \cite{Watanabe:2010fh}.
The same conclusion also applies to more general anisotropic 
backgrounds \cite{Hervik}.
There remains the rotational symmetry in the $(y,z)$ plane. 
Hence, we can take the metric ansatz 
\begin{equation}
ds^2 = -dt^2 + e^{2\alpha(t)} \left[ e^{-4\sigma (t)}dx^2
+e^{2\sigma (t)}(dy^2+dz^2) \right] \ ,
\label{anisotropic-metric}
\end{equation}
where $e^\alpha \equiv a$ and $\sigma$ are the isotropic 
scale factor and the spatial shear, respectively.
It is easy to solve the equation of motion for $v$ as
\begin{equation}
\dot{v} =p_A\,f(\phi)^{-2}  e^{-\alpha -4\sigma}\,,
\label{eq:Ax}
\end{equation}
where $p_A$ is a constant of integration and an overdot denotes 
a derivative with respect to the cosmic time $t$. 

The Friedmann equation and the inflaton equation of motion are 
given, respectively,  by 
\begin{eqnarray}
H^2 &=& \dot{\sigma}^2+\frac{1}{3 M_{\rm pl}^2} \left[ \frac{1}{2}\dot{\phi}^2
+V(\phi )+\frac{p_A^2}{2}f(\phi )^{-2}e^{-4\alpha -4\sigma} \right] \,, 
\label{eq:hamiltonian-vector} \\
\ddot{\phi} &=& -3\dot{\alpha}\dot{\phi}-V_{,\phi}
+p_A^2f(\phi)^{-3} f_{,\phi}e^{-4\alpha -4\sigma}\,,
\label{inflatoneq}
\end{eqnarray}
where $H\equiv \dot{\alpha}$ is the Hubble expansion rate, 
and $V_{,\phi} \equiv dV/d\phi$, $f_{,\phi} \equiv df/d\phi$. 
We define the energy density of the vector field as
\begin{equation}
\rho_A \equiv \frac{p_A^2}{2} f(\phi)^{-2}e^{-4\alpha -4\sigma} \,.
\label{rhoA}
\end{equation} 
In order to sustain inflation, the potential energy $V(\phi)$
of the inflaton needs to dominate over $\dot{\phi}^2/2$ and $\rho_A$.
Since the shear term $\Sigma \equiv \dot{\sigma}$ should be suppressed 
relative to $H$, the Friedmann equation (\ref{eq:hamiltonian-vector}) reads
\begin{equation}
H^2 =\dot{\alpha}^2 \simeq \frac{V(\phi)}{3M_{\rm pl}^2}\,.
\label{Friap}
\end{equation}

If $f$ is a rapidly decreasing function in time, 
it happens that $\rho_A$ does not decay.
In particular, for the coupling 
\begin{equation}
f(\phi)=e^{-2\alpha}=a^{-2}\,,
\label{fphi1}
\end{equation}
the energy density (\ref{rhoA}) stays nearly constant 
(under the approximation $|\sigma| \ll \alpha$).
In this case, neglecting the contribution of 
the vector field on the r.h.s. of Eq.~(\ref{inflatoneq}), 
the field $\phi$ satisfies the slow-roll equation of motion
$3\dot{\alpha}\dot{\phi} \simeq -V_{,\phi}$. 
Combining this equation with Eq.~(\ref{Friap}), 
it follows that $d\alpha/d\phi \simeq -V/(M_{\rm pl}^2V_{,\phi})$.
Then, the critical coupling (\ref{fphi1}) can be expressed as
\begin{equation}
f(\phi)=e^{2\int \frac{V}{M_{\rm pl}^2 V_{,\phi}}
d\phi }\,.
\label{fphive}
\end{equation}
The shear term obeys the equation of motion 
\begin{equation}
\dot{\Sigma}=-3H \Sigma+\frac{2\rho_A}{3M_{\rm pl}^2}\,.
\end{equation}
If an anisotropy converges to a nearly constant value $\Sigma$, 
then the quantity $\Sigma/H$ reduces to 
\begin{equation}
\frac{\Sigma}{H} \simeq \frac{2\rho_A}{3V}\,,
\label{sigvector}
\end{equation}
where we used Eq.~(\ref{Friap}). 
In the following, let us estimate the ratio $\Sigma/H$ during 
anisotropic inflation.
Ignoring the $\ddot{\phi}$ term in Eq.~(\ref{inflatoneq}) and 
combining it with Eq.~(\ref{Friap}), it follows that 
\begin{equation}
\frac{d\phi}{d\alpha} \simeq -\frac{M_{\rm pl}^2V_{,\phi}}{V}
+\frac{2p_A^2}{V_{,\phi}}e^{-4\alpha-4\sigma-4\int \frac{V}
{M_{\rm pl}^2 V_{,\phi}}d\phi}\,,
\label{dphial}
\end{equation}
where we used Eq.~(\ref{fphive}).
Neglecting the variation of $\phi/M_{\rm pl}$ relative to 
that of $\alpha$, we can integrate Eq.~(\ref{dphial}) to give
\begin{equation}
e^{4\alpha+4\sigma+4\int \frac{V}
{M_{\rm pl}^2 V_{,\phi}}d\phi} \simeq 
\frac{8p_A^2 V}{M_{\rm pl}^2 V_{,\phi}^2} 
(\alpha+\alpha_0)\,,
\label{intvec}
\end{equation}
where $\alpha_0>0$ is an integration constant. 
Substituting Eq.~(\ref{intvec}) into Eq.~(\ref{rhoA}), we have
\begin{equation}
r_A \equiv \frac{\rho_A}{\epsilon V} \simeq \frac{1}{8(\alpha+\alpha_0)}\,,
\label{rA}
\end{equation}
where $\epsilon \equiv -\dot{H}/H^2$, and 
we used the fact that $\epsilon \simeq 
(M_{\rm pl}^2/2)(V_{,\phi}/V)^2$ under the slow-roll approximation.
Provided that $\alpha \ll \alpha_0$, the ratio $r_A$ is nearly constant.
As we will see in Sec.~\ref{perksec}, the quantity $r_A$ 
is related to the anisotropic parameter $g_*$ appearing in the
power spectrum of curvature perturbations. 
{}From Eqs.~(\ref{sigvector}) and (\ref{rA}) we obtain 
\begin{equation}
\frac{\Sigma}{H} \simeq \frac{1}{12(\alpha+\alpha_0)}\epsilon\,,
\label{SigH1}
\end{equation}
which means that the anisotropy survives during inflation 
for $\alpha \ll \alpha_0$.

The above discussion can be generalized to the coupling 
of the form 
\begin{equation}
f(\phi)=e^{2c\int \frac{V}{M_{\rm pl}^2 V_{,\phi}}
d\phi }\,,
\label{geneco}
\end{equation}
where $c$ is a constant parameter. 
If the condition 
\begin{equation}
c=\frac{M_{\rm pl}^2}{2} \frac{f_{,\phi}}{f} \frac{V_{,\phi}}{V}>1
\end{equation}
is satisfied, the energy density of the vector field grows as 
$\rho_A \propto e^{4(c-1)\alpha}$ during the slow-roll phase 
of the inflaton. Eventually, the vector field becomes relevant to 
the inflaton dynamics governed by Eq.~(\ref{inflatoneq}).
However, when the third term on the r.h.s. of Eq.~(\ref{inflatoneq})
dominates over the second term, the inflaton does not roll down, 
which makes $\rho_A $ decrease. 
Hence the condition $\rho_A \ll V(\phi)$ is always satisfied.
In this way, there appears an attractor where inflation continues even 
when the vector field affects the inflaton dynamics.
For the coupling (\ref{geneco}) the shear to the Hubble expansion 
rate approaches the value \cite{Watanabe}
\begin{equation}
\frac{\Sigma}{H} \simeq \frac13 \frac{c-1}{c}\epsilon\,.
\label{SigHvector}
\end{equation}
Thus, inflation is slightly anisotropic and the energy density of 
the vector field never decays for the coupling (\ref{geneco}) with $c>1$.
From Eqs.~(\ref{sigvector}) and (\ref{SigHvector}) it follows that the ratio $r_A$ 
defined in Eq.~(\ref{rA}) is nearly constant, i.e., 
$r_A \simeq (c-1)/(2c)$. 

In the attractor regime where the vector field contributes 
to the dynamics of the system there is the relation
$d\alpha/d\phi \simeq -cV/(M_{\rm pl}^2V_{,\phi})$ \cite{review2}, 
in which case the coupling (\ref{geneco}) evolves as $f(\phi) \propto a^{-2}$, 
i.e., the same as Eq.~(\ref{fphi1}).
In Sec.~\ref{perksec} we shall use this property for the evaluation 
of two-point correlation functions of primordial perturbations.

%%%%%
\subsection{$f(\phi)^2H_{\mu \nu \lambda}H^{\mu \nu \lambda}$ model}
\label{twoformsec}
%%%%%

In the presence of a two-form field coupled to the inflaton \cite{Ohashi}, 
anisotropic hair can survive during inflation as in the vector model. 
In this case, the action reads
\begin{equation}
S = \int d^4 x \sqrt{-g} \left[\frac{M_{\rm pl}^2}{2}R 
- \frac{1}{2}\partial_\mu \phi \partial^\mu \phi - V(\phi )
-\frac{1}{12}f(\phi)^2 H_{\mu\nu\lambda} 
H^{\mu\nu\lambda} \right]\,,
\label{oriaction}
\end{equation}
where the field strength  $H_{\mu\nu\lambda}$ is related to 
a two-form field $B_{\mu \nu}$, as 
\begin{equation}
H_{\mu\nu \lambda} = 
\partial_\mu B_{\nu \lambda} + \partial_\nu B_{\lambda\mu}
+ \partial_\lambda B_{\mu\nu}\,.
\label{Hdef}
\end{equation}
Without loss of generality, one can take the $(y,z)$ plane 
in the direction of the two-form field. 
Then we can express $B_{\mu \nu}$ in the form
\begin{equation}
\frac{1}{2}B_{\mu\nu}\,dx^{\mu}\wedge dx^{\nu}= 
w(t)\,dy \wedge dz\,,  
\end{equation}
where $w(t)$ is a function with respect to $t$.
Since there exists a rotational symmetry in the $(y,z)$ plane, 
the metric can be parametrized by the form (\ref{anisotropic-metric}).
The equation of motion for the two-form field $w$ 
is easily solved as
\begin{equation}
\dot{w} = p_B\,f(\phi)^{-2}  e^{\alpha +4\sigma}\,,
\label{eq:Ax2}
\end{equation}
where $p_B$ is a constant of integration. 

The Hubble parameter $H=\dot{\alpha}$ and the inflaton $\phi$ 
obeys the equations of motion 
\begin{eqnarray}
H^2 &=& \dot{\sigma}^2+\frac{1}{3M_{\rm pl}^2} 
\left[ \frac{1}{2}\dot{\phi}^2+V(\phi )
+\frac{p_B^2}{2}f(\phi )^{-2}e^{-2\alpha + 4\sigma} \right]\,,
\label{eq:hamiltonian}\\
\ddot{\phi} &=& -3\dot{\alpha}\dot{\phi}-V_{,\phi}
+p_B^2f(\phi)^{-3} f_{,\phi}e^{-2\alpha + 4\sigma}\,,
\label{inflatoneq2}
\end{eqnarray}
which are analogous to Eqs.~(\ref{eq:hamiltonian-vector}) and 
(\ref{inflatoneq}) with the difference of the exponential factors.
Defining the energy density of the two-form field as
\begin{equation}
\label{rho-v-section-4}
\rho_B \equiv \frac{p_B^2}{2}f(\phi)^{-2}e^{-2\alpha +4\sigma}\,,
\end{equation}
it follows that $\rho_B$ stays nearly constant for the coupling 
\begin{equation}
f(\phi)=e^{-\alpha}=a^{-1}\,.
\label{couplingtwo}
\end{equation}
In the slow-roll regime of the inflaton there is the relation 
$d\alpha/d\phi \simeq -V/(M_{\rm pl}^2 V_{,\phi})$, 
so that the coupling (\ref{couplingtwo}) can be expressed as
\begin{equation}
f(\phi)=e^{\int \frac{V}{M_{\rm pl}^2 V_{,\phi}}d\phi}\,.
\label{coutwo}
\end{equation}
Since the shear $\Sigma=\dot{\sigma}$ satisfies the equation of motion 
\begin{equation}
\dot{\Sigma}=-3H \Sigma-\frac{2\rho_B}{3M_{\rm pl}^2}\,,
\end{equation}
the ratio $\Sigma/H$ should converge to
\begin{equation}
\frac{\Sigma}{H}=-\frac{2\rho_B}{3V}\,,
\label{sigtwo}
\end{equation}
where Eq.~(\ref{Friap}) is used. 
The sign of $\Sigma/H$ is opposite to that of the vector model. 
During anisotropic inflation the ratio (\ref{sigtwo}) reads \cite{Ohashi}
\begin{equation}
\frac{\Sigma}{H} \simeq -\frac{1}{3(\alpha+\alpha_0)}\epsilon\,,
\label{sigtwo2}
\end{equation}
where $\alpha_0>0$ is a constant.
{}From Eqs.~(\ref{sigtwo}) and (\ref{sigtwo2}) we have 
\begin{equation}
r_B \equiv \frac{\rho_B}{\epsilon V} \simeq 
\frac{1}{2(\alpha+\alpha_0)}\,,
\label{rB}
\end{equation}
which is nearly constant for $\alpha \ll \alpha_0$.
The ratio (\ref{rB}) appears in the anisotropic scalar 
power spectrum.

We can generalize the coupling (\ref{coutwo}) to the form 
\begin{equation}
f(\phi)=e^{c\int \frac{V}{M_{\rm pl}^2 V_{,\phi}}d\phi }\,,
\label{coutwo2}
\end{equation}
where $c$ is a constant.
For the super-critical case characterized by 
\begin{equation}
c=M_{\rm pl}^2 \frac{f_{,\phi}}{f} \frac{V_{,\phi}}{V}>1\,,
\end{equation}
there is an attractor solution along which the ratio 
$\Sigma/H$ approaches the value \cite{Ohashi}
\begin{equation}
\frac{\Sigma}{H} \simeq -\frac23 \frac{c-1}{c} \epsilon\,,
\label{SigHtwo}
\end{equation}
whose sign is opposite to Eq.~(\ref{SigHvector}).
In this regime there is the relation 
$d\alpha/d\phi \simeq -cV/(M_{\rm pl}^2 V_{,\phi})$, 
so that the coupling (\ref{coutwo2}) evolves as Eq.~(\ref{couplingtwo}).
Note that the ratio $r_B$ is nearly constant, i.e., 
$r_B \simeq (c-1)/c$.

The anisotropy induced by the two-form field is the prolate-type,  
in contrast to the vector field which induces the oblate-type anisotropy. 
This difference comes from the fact the vector $A_{\mu}$ 
extending to the $x$-direction speeds down the expansion 
in that direction, while the two-form field $B_{\mu \nu}$
extending in the $(y,z)$ plane speeds down the expansion
in the $(y,z)$-direction. 

We note that the couplings (\ref{fphive}) and (\ref{coutwo}), which 
give rise to anisotropic inflation, are present for any slow-roll 
inflaton potentials. For the exponential potential $V(\phi)=
V_0e^{\lambda \phi/M_{\rm pl}}$ the couplings $f(\phi)$ are 
of the exponential forms $f(\phi) \propto e^{\mu \phi/M_{\rm pl}}$ \cite{powerlaw}, 
as they often appear as a dilatonic coupling in string theory. 
For some inflaton potentials the functions $f(\phi)$ may not 
be so natural, but there is a possibility that such couplings 
can be motivated by future development of 
string theory or supergravity. 
It is worth mentioning that power-law kinetically driven 
anisotropic inflation (k-inflation \cite{kinf}) can be generally realized
for the exponential couplings 
$f(\phi) \propto e^{\mu \phi/M_{\rm pl}}$ \cite{anikinf}.

%===============================================================%
%************************ SECTION III ****************************%
%===============================================================%

%%%%%%%%%%%%%%%%%%%%%%%%%%%%%%%%%
\section{Scalar and tensor power spectra and their correlations}
\label{perksec}
%%%%%%%%%%%%%%%%%%%%%%%%%%%%%%%%%

In order to study observational signatures of anisotropic inflation, 
we need to know the two-point correlation functions of curvature 
perturbations and gravitational waves as well as their cross correlations.
For the vector model the power spectrum of curvature perturbations 
was derived in Refs.~\cite{Gum,Watanabe:2010fh,Bartolo}, whereas the anisotropic contribution 
to gravitational waves in the same model was discussed 
in Refs.~\cite{Gum,Watanabe:2010fh}. 
For the two-form field model the present authors obtained the 
anisotropic scalar power spectrum \cite{Ohashi}, but the 
tensor power spectrum has not been derived yet.
We also note that in the vector model the correlation between the 
temperature perturbation and the B-mode polarization was 
studied in Ref.~\cite{Watanabe2}, 
but the cross-correlation between scalar and tensor perturbations
in the two-form model has not been studied.
Here we provide all the formulas of these observables convenient 
to confront with observations.
 
Since the anisotropy of the expansion rate needs to be sufficiently small 
for the compatibility with observations, it is a good approximation to 
neglect the effect of the anisotropic expansion for the derivation of 
the perturbation equations \cite{Watanabe:2010fh}.  
The effect of the anisotropy appears in the interacting Hamiltonians 
between vector/two-form fields and scalar/tensor perturbations, 
by which the scalar/tensor power spectra are modified.
Then, we consider a general perturbed metric with four scalar functions
$A,B,\psi,E$ and the tensor perturbation $h_{ij}$ about the 
flat Friedmann-Lema\^itle-Robertson-Walker (FLRW) 
background \cite{Bardeen}:
\begin{equation}
ds^2 = a(\tau)^2 \left\{-(1+2A)d\tau^2 +2 \partial_i B d\tau dx^i
+[(1+2\psi)\delta_{ij}+2\partial_{ij}E+h_{ij}]dx^i dx^j \right\}\,,
\label{permet}
\end{equation}
where $\tau=\int a^{-1} dt$ is the conformal time.
After the end of inflation, the coupling $f(\phi)$ approaches a constant because
the inflaton stabilizes at the potential minimum. In this case vector perturbations 
decay after inflation as in the standard scenario, so 
we neglect its contribution to the CMB observables 
relative to those of scalar and metric perturbations.
We introduce the gauge-invariant comoving 
curvature perturbation \cite{Bardeen2} 
(see also Refs.~\cite{zetadef}):
\begin{equation}
\zeta=\psi-\frac{H}{\dot{\phi}}\delta \phi\,,
\end{equation}
where $\delta \phi$ is the perturbation of the inflaton $\phi$.
In the following we choose the spatially flat gauge ($\psi=0$), 
in which case $\zeta=-(H/\dot{\phi})\delta \phi$. 
The curvature perturbation can be expressed in terms of 
the Fourier components with the comoving 
wave number ${\bm k}$, as
\begin{equation}
\zeta({\bm x}, \tau)=\int \frac{d^3 k}{(2\pi)^{3/2}}
e^{i {\bm k}\cdot {\bm x}} \hat{\zeta}({\bm k},\tau)\,,\qquad
\hat{\zeta}({\bm k},\tau)=\zeta(k,\tau)a({\bm k})
+\zeta^{*}(k,\tau) a^\dagger(-{\bm k})\,,
\label{zeta0def}
\end{equation}
where the annihilation and creation operators $a({\bm k})$ 
and $a^\dagger({\bm k'})$ satisfy the commutation relation 
$[a({\bm k}), a^\dagger({\bm k'}) ]=\delta^{(3)} ({\bm k}-{\bm k}')$.
We define the scalar power spectrum ${\cal P}_{\zeta}$ 
in terms of the two-point correlation function of $\zeta$, as
\begin{equation}
\langle \hat{\zeta}({\bm k}_1) \hat{\zeta}({\bm k}_2) \rangle
=\frac{2\pi^2}{k_1^3} \delta^{(3)}({\bm k}_1+{\bm k}_2) 
{\cal P}_{\zeta}(k_1)\,.
\end{equation}

We decompose $\zeta$ into the isotropic field $\zeta^{(0)}$ 
and the contribution $\delta \zeta$ coming from the anisotropic fields, as
\begin{equation}
\zeta=\zeta^{(0)}+\delta \zeta\,.
\end{equation}
In what follows we shall focus on the couplings 
(\ref{fphive}) and (\ref{coutwo}), i.e., $c=1$. 
The situation is similar for the general couplings 
(\ref{geneco}) and (\ref{coutwo2}) with $c$ close to 1.
Then we can employ the usual slow-roll relations 
$\dot{\phi}/H \simeq -M_{\rm pl}^2 V_{,\phi}/V$ and  
$\epsilon=-\dot{H}/H^2 \simeq (M_{\rm pl}^2/2)(V_{,\phi}/V)^2$, 
so that $\zeta^{(0)} \simeq \delta \phi/(M_{\rm pl}\sqrt{2\epsilon})$.
The solution to the Fourier mode $\zeta^{(0)}(k, \tau)$, which recovers 
the Bunch-Davies vacuum state for the field perturbation 
$\delta \phi$ in the asymptotic past ($k \tau \to -\infty$), is 
given by \cite{Maldacena:2002vr}
\begin{equation}
\zeta^{(0)}(k,\tau)=\frac{H(1+ik\tau)}
{2\sqrt{\epsilon}M_{\rm pl} k^{3/2}}
e^{-ik \tau}\,.
\label{backsol}
\end{equation}
The power spectrum can be written as the sum of the two contributions from 
$\zeta^{(0)}$ and $\delta \zeta$, as ${\cal P}_{\zeta}={\cal P}^{(0)}_{\zeta}+\delta {\cal P}_{\zeta}$.
Using the solution (\ref{backsol}) long time after the Hubble radius
crossing ($\tau \to 0$), the isotropic power spectrum of $\zeta$ 
is given by 
\begin{equation}
{\cal P}_{\zeta}^{(0)}=\frac{H^2}{8\pi^2 \epsilon M_{\rm pl}^2}\,.
\label{powerli}
\end{equation}
In Secs.~\ref{pervec} and \ref{pertwo} we shall evaluate the anisotropic 
corrections to ${\cal P}_{\zeta}^{(0)}$ in both the vector and 
the two-form field models.

For the tensor perturbation $h_{ij}$ we impose the traceless and 
transverse conditions $h_{ii}=h_{ij,j}=0$, as usual.  
The second-order action for $h_{ij}$ reads
\begin{equation}
S_h = \frac{M_{\rm pl}^2}{4} \int d\tau d^3x\, a^2 \left[ 
\frac{1}{2} h'_{ij} h'_{ij} - \frac{1}{2} h_{ij,k}h_{ij,k} \right]\,,
\end{equation}
where the prime denotes the differentiation 
with respect to $\tau$.  
We have two physical degrees of freedom for $h_{ij}$ which
can be characterized by the symmetric polarization tensors
$e^{(+,\times)}_{ij}(\bm{k})$ satisfying 
\begin{equation}
e^{(s)}_{ii} (\bm{k}) = 0 \ , \qquad k_j  
e^{(s)}_{ij} (\bm{k}) = 0\,,
\label{pore1}
\end{equation}
where $s= +, \times $ represent the polarizations. 
It is convenient to adopt the normalization
\begin{equation}
e^{(s)}_{ij} (\bm{k})  e^{*(s')}_{ij} (\bm{k}) = \delta_{ss'} \,,
\label{basis_prop1}
\end{equation}
where $*$ represents a complex conjugate.
Remark that the following relation holds:
\begin{eqnarray}
e^{(s)}_{ij} (\bm{k}) = e^{*(s)}_{ij} (-\bm{k})\,.
\label{basis_prop2}
\end{eqnarray}

Now, it is straightforward to quantize tensor perturbations.
The mode expansion can be written as \cite{Soda:2011am}
\begin{equation}
h_{ij} (\bm{x},\tau) = \int \frac{d^3k}{(2\pi)^{3/2}}
e^{i {\bm k}\cdot {\bm x}} \hat{h}_{ij}({\bm k},\tau)\,,\qquad
\hat{h}_{ij}({\bm k},\tau)=
\sum_{s= +,\times} \left[ 
h_s (k,\tau)a_s ({\bm k})+h_s^*(k,\tau)
a_s^{\dagger} (-{\bm k}) \right]e_{ij}^{(s)}({\bm k})\,,
\label{hijde}
\end{equation}
where the creation and annihilation operators are normalized as 
$ \left[ a_s (\bm{k}) , a^\dagger_s (\bm{k}') \right] 
=\delta_{ss'} \delta^{(3)} (\bm{k}-\bm{k}' )$.
We define the tensor power spectrum ${\cal P}_{h}$, as 
\begin{equation}
\langle \hat{h}_{ij}({{\bm k}_1}) \hat{h}_{ij}({{\bm k}_2}) \rangle
=\frac{2\pi^2}{k_1^3} \delta^{(3)}({\bm k}_1+{\bm k}_2) 
{\cal P}_h(k_1)\,.
\label{tensorspe}
\end{equation}
When we study the polarization of tensor perturbations, 
we can take both the vectors $\bm k_1$ and $\bm k_2$ 
lying on the ($x, y$)-plane without lose of 
generality (because of the momentum 
conservation $\bm{k}_1 + \bm{k}_2 =0$).
In this case we can take
\begin{equation}
\bm{k}_1 = k_1 \left( \cos \theta , \sin\theta , 0 \right) \,,
\end{equation}
where $\theta$ represents the angle between $\bm k_1$ and $x$-axis.
For $\bm{k}_1=(k_1, 0, 0)$, i.e., $\theta=0$,  
the polarization tensors $e^{(s)}_{ij} (\bm{k}_1)$
satisfying the relations (\ref{pore1})-(\ref{basis_prop2}) are  
\begin{eqnarray}
e^{(+)}_{ij} (\bm{k}_1) 
= \frac{1}{\sqrt{2}}\left( 
\begin{array}{ccc}
 0 & 0 & 0 \\
 0 & 1 & 0 \\
 0 & 0 & -1
\end{array}
\right)\,,\qquad
e^{(\times)}_{ij}  (\bm{k}_1) 
= \frac{i}{\sqrt{2}}\left( 
\begin{array}{ccc}
 0 & 0 & 0 \\
 0 & 0 &  1 \\
 0 &  1 & 0
\end{array}
\right) \,.
\end{eqnarray}
To obtain the polarization for $\bm{k}_1=k_1(\cos \theta , \sin\theta , 0)$, 
we need to rotate the above one by $\theta$ as
\begin{eqnarray}
e^{(+)}_{ij}  (\bm{k}_1) 
= \frac{1}{\sqrt{2}}\left( 
\begin{array}{ccc}
 \sin^2 \theta & -\sin \theta \cos\theta & 0 \\
 -\sin\theta \cos\theta & \cos^2 \theta & 0 \\
 0  & 0 & -1
\end{array}
\right)\,,\qquad
e^{(\times)}_{ij}  (\bm{k}_1) 
= \frac{i}{\sqrt{2}}\left( 
\begin{array}{ccc}
 0 & 0 & - \sin\theta \\
 0 & 0 &  \cos\theta \\
 - \sin \theta  &  \cos \theta & 0
\end{array}
\right)\,.
\label{eij}
\end{eqnarray}
We write the Fourier mode $\hat{h}_{ij}({\bm k}, \tau)$ in Eq.~(\ref{hijde}), as 
\begin{equation}
\hat{h}_{ij} ({\bm k},\tau)=\sum_{s= +,\times} 
\hat{h}_s ({\bm k},\tau) e_{ij}^{(s)} ({\bm k})\,,\qquad
\hat{h}_s ({\bm k}, \tau)=h_s (k,\tau)a_s ({\bm k})+h_s^*(k,\tau)
a_s^{\dagger} (-{\bm k})\,.
\label{hijbmk}
\end{equation}
Using Eq.~(\ref{eij}), it follows that 
\begin{equation}
\hat{h}_{ij} ({\bm k}_1, \tau)=\frac{1}{\sqrt{2}}\left(
\begin{array}{ccc}
\hat{h}_+\sin^2\theta & -\hat{h}_+\sin\theta\cos\theta & -i \hat{h}_{\times}\sin\theta \\
-\hat{h}_+\sin\theta\cos\theta & \hat{h}_+ \cos^2\theta & i \hat{h}_{\times}\cos\theta \\
-i \hat{h}_{\times}\sin\theta & i \hat{h}_{\times}\cos\theta & -\hat{h}_+ 
\end{array}\right)\,,
\label{delh}
\end{equation}
which will be used for the evaluation of the interacting Hamiltonians 
between gravitational waves and vector/two-form fields.

We decompose the tensor perturbation $\hat{h}_{ij}$ into 
the isotropic field $\hat{h}_{ij}^{(0)}$ and the anisotropic 
contribution $\delta \hat{h}_{ij}$.
The isotropic mode function $u_k^{(0)} \equiv M_{\rm pl} \sqrt{k/2}\,h_s^{(0)}(k)$ 
obeys the following evolution equation
\begin{equation}
{u_k^{(0)}}'' + 2 \frac{a'}{a} {u_k^{(0)}}' + k^2 u_k^{(0)} = 0\,,
\end{equation}
where the canonical commutation relation leads to the
normalization condition
\begin{eqnarray}
u^{*(0)}_k {u_k^{(0)}}^{'}
- u_k^{(0)} {u^{*(0)}_k}^{'}  = -\frac{2ik}{a^2} \ .
\label{modefunction:normalization}
\end{eqnarray}
Once a set of  mode functions satisfying this normalization is
specified,  the corresponding Fock vacuum is determined 
by $a_s(\bm{k})|0\rangle=0$. 
The mode function in a de Sitter background 
is given by $u_k^{(0)}(\tau) = (H/k) \left( 1 + ik\tau \right) e^{-ik\tau}$, 
that is  
\begin{equation}
h_s^{(0)}(k,\tau)= \frac{\sqrt{2}H}{M_{\rm pl}k^{3/2}} 
\left( 1 + ik\tau \right) e^{-ik\tau } \qquad (s=+,\times).
\label{hsso}
\end{equation}
Using this solution and (\ref{hijbmk}), (\ref{delh}) long after the Hubble radius crossing, 
the isotropic power spectrum defined by (\ref{tensorspe}) reads
\begin{equation}
{\cal P}_h^{(0)}=\frac{2H^2}{\pi^2 M_{\rm pl}^2}
=16\epsilon {\cal P}_{\zeta}^{(0)}\,.
\label{Pten}
\end{equation}
In the following we evaluate the anisotropic corrections to ${\cal P}_{h}^{(0)}$ 
in both the vector and the two-form models.
In doing so, it is convenient to notice the following commutation relations 
\begin{eqnarray}
&&[\hat{\zeta}^{(0)}({\bm k},\tau), \hat{\zeta}^{(0)}({\bm k'},\tau')]
\simeq -i \frac{H^2}{6\epsilon M_{\rm pl}^2} (\tau^3-\tau{'^3})
\delta^{(3)} ({\bm k}+{\bm k}')\,, \label{commutator1}  \\
&&[\hat{h}_s^{(0)}({\bm k}, \tau) , \hat{h}_s^{(0)}({\bm k}^{\prime} , \tau^{\prime})]
\simeq -i\frac{4H^2}{3M_{\rm pl}^2}(\tau^3-{\tau^{\prime}}^3)\delta^{(3)} ({\bm k} + {\bm k}^{\prime})\,, \label{commutator3} 
\end{eqnarray}
which can be derived by employing the solutions 
(\ref{backsol}) and (\ref{hsso}) in the super-Hubble regime ($|k \tau| \ll 1$).

%%%%%%%%%%%%%%%%%%%%%%%%%%%%
\subsection{$f(\phi)^2 F_{\mu \nu}F^{\mu \nu}$ model}
\label{pervec}
%%%%%%%%%%%%%%%%%%%%%%%%%%%%

For the model described by the action (\ref{eq:action})
we decompose the vector field $A_{\mu}$
into the Fourier components by choosing the Coulomb gauge:
\begin{equation}
A_i({\bm x},\tau) =A_i^{(0)} (\tau)+ \delta A_i 
= A_i^{(0)} (\tau)+
\sum_{\lambda=1,2} \int \frac{d^3k}{(2\pi)^{3/2}} 
e^{i{\bm k} \cdot {\bm x}} \left[ 
A_{\lambda}(k, \tau)a_{\lambda}({\bm k}) 
+A_{\lambda}^* (k,\tau)a_{\lambda}^{\dagger} (-{\bm k}) \right]
{\epsilon}^{(\lambda)}_i ({\bm k})\,,
\end{equation}
where $A_i^{(0)} (\tau)=(A_x^{(0)},0,0)$ is the background component, and 
${\epsilon}^{(\lambda)}_i ({\bm k})$ ($\lambda=1,2$) 
are polarization vectors satisfying the relations
\begin{equation}
k_i {\epsilon}^{(\lambda)}_i ({\bm k}) =0 \,, \quad 
{\epsilon}^{(\lambda)}_i(-{\bm k}) ={ \epsilon}_i^{{*(\lambda)}}({\bm k}) \,, \quad
{ \epsilon}^{(\lambda)}_i ({\bm k}) \  { \epsilon}^{*{(\lambda ')}}_i ({\bm k})
=\delta_{\lambda\lambda '} \,.
\end{equation}
With the previous parametrization $k_i = k (\cos\theta ,\sin\theta ,0)$, 
an explicit representation is given by
\begin{equation}
\epsilon^{(1)}_i = (-i\sin \theta , i\cos \theta , 0) \ , 
\qquad \epsilon^{(2)}_i = (0,0,1)  \ .
\end{equation}

The rescaled field $V_{\lambda}=fA_{\lambda}$ obeys 
the equation of motion 
\begin{equation}
V_{\lambda}''+\left(k^2 -\frac{f''}{f} \right)V_{\lambda}=0\,.
\label{Vlam}
\end{equation}
For the coupling $f$ given by Eq.~(\ref{fphi1}), we have 
$f \propto \tau^2$ on the de Sitter background ($a=-(\tau H)^{-1}$) and hence 
$f''/f=2/\tau^2$. In this case the resulting vector field perturbation 
is scale-invariant. 
The solution to Eq.~(\ref{Vlam}), which recovers 
the Bunch-Davies vacuum in the asymptotic past, is
\begin{equation}
A_{\lambda}(k, \tau)= 
\frac{Ha^3}{\sqrt{2k^3}}(1+ik \tau)e^{-i k\tau}\,.
\label{Alam}
\end{equation}
It is convenient to define the electric components
\begin{equation}
E_x \equiv \frac{f}{a^2}{A_x^{(0)}}' \,, \qquad 
\delta E_i \equiv \frac{f}{a^2} \delta A_i' \,,
\label{Edef}
\end{equation}
where $E_x$ and $\delta E_{i}$ correspond to the background 
and the perturbed values.

The next step is to derive anisotropic contributions $\delta{\cal P}_{\zeta}$ 
and $\delta{\cal P}_h$ to the isotropic scalar and tensor power 
spectra (\ref{powerli}) and (\ref{Pten}).
The tree-level interacting Lagrangian is
\begin{eqnarray}
L_{\rm int}&=&-\frac{a^4}{4}\left(\langle f^2 \rangle 
+\frac{\partial \langle f^2 \rangle}{\partial \phi}\delta \phi\right)
(\langle F_{\mu\nu} \rangle+\delta F_{\mu\nu})
(\langle F^{\mu\nu} \rangle+\delta F^{\mu\nu})  \nonumber \\
&\simeq& f^2{A_x^{(0)}}' \left(4 \delta A_x^{\prime} \zeta - \delta A_x^{\prime} h_{xx} 
-  \delta A_y^{\prime} h_{xy} - \delta A_z^{\prime} h_{xz} \right) \nonumber \\
&=& a^4 E_x \left(4 \delta E_x \zeta - \delta E_x h_{xx} - \delta E_y h_{xy} 
- \delta E_z h_{xz} \right) \,,
\label{Lint_vec}
\end{eqnarray}
where in the first line $\langle \, \rangle$ represents the background value and 
after the second line we picked up the second-order perturbation terms 
and dropped the symbol $\langle \, \rangle$.
We also used $(\partial \langle f^2 \rangle/\partial \phi)\delta \phi=4f^2\zeta$, 
which follows from the relation $\zeta=-(H/\dot{\phi})\delta \phi$ and the 
slow-roll conditions.

We decompose $\delta E_i ({\bm x}, \tau)$ into the Fourier components, as
\begin{equation}
\delta E_i ({\bm x}, \tau)=\int \frac{d^3k}{(2\pi)^{3/2}} e^{i {\bm k} \cdot {\bm x}}
\delta {\cal E}_i ({\bm k}, \tau)\,.
\label{delEi}
\end{equation}
Using the solution (\ref{Alam}) in the super-Hubble regime $|k \tau| \ll 1$, 
the mode function $\delta {\cal E}_i ({\bm k}, \tau)$ is given by 
\begin{equation}
\delta {\cal E}_i ({\bm k}, \tau)=
\sum_{\lambda=1,2} \frac{3H^2}{\sqrt{2k^3}} 
\left[ a_{\lambda} ({\bm k})+a_{\lambda}^{\dagger} (-{\bm k}) 
\right] {\epsilon}^{(\lambda)}_i ({\bm k})\,.
\end{equation}
The contributions to the interacting Hamiltonian 
$H_{\rm int}=- \int d^3x\,L_{\rm int}$, which come from 
the four interacting Lagrangians in Eq.~(\ref{Lint_vec}), 
are given, respectively, by
\begin{eqnarray}
H_{\zeta}&=&-\frac{4E_x}{H^4 \tau^4} \int d^3k\, 
\delta {\cal E}_x({\bm k}, \tau) \hat{\zeta}^{(0)}(-{\bm k}, \tau) \,, \label{Hzeta} \\
H_{h1}&=&\frac{E_x}{\sqrt{2}H^4 \tau^4} \int d^3k\, \delta {\cal E}_x({\bm k}, \tau) 
\hat{h}_+^{(0)}(-{\bm k}, \tau) \sin^2\theta_{{\bm k}, {\bm x}} \,, \\
H_{h2}&=&-\frac{E_x}{\sqrt{2}H^4 \tau^4} \int d^3k\, \delta {\cal E}_y({\bm k}, \tau) 
\hat{h}_+^{(0)}(-{\bm k}, \tau) \sin\theta_{{\bm k}, {\bm x}} \cos\theta_{{\bm k}, {\bm x}}  \,, \\
H_{h3}&=&\frac{i E_x}{\sqrt{2}H^4 \tau^4} \int d^3k\, \delta {\cal E}_z({\bm k}, \tau) 
\hat{h}_{\times}^{(0)}(-{\bm k}, \tau) \sin\theta_{{\bm k}, {\bm x}} \,,
\label{Hint_vec1}
\end{eqnarray}
where $\theta_{{\bm k},{\bm x}}$ is the angle between the wave number 
${\bm k}$ and the $x$-axis.
In deriving the above Hamiltonians, 
we used Eqs.~(\ref{zeta0def}), (\ref{hijde}), (\ref{delh}), (\ref{delEi}), 
and replaced $\hat{\zeta}(-{\bm k}, \tau)$ and $\hat{h}_{s}(-{\bm k}, \tau)$
for the isotropic perturbations $\hat{\zeta}^{(0)}(-{\bm k}, \tau)$ and 
$\hat{h}^{(0)}_{s}(-{\bm k}, \tau)$, respectively.

The two-point correction of scalar perturbations following from 
the interacting Hamiltonian (\ref{Hzeta}) to the isotropic 
power spectrum ${\cal P}_{\zeta}^{(0)}$ reads
\begin{eqnarray}
\delta \langle 0|\hat{\zeta}({\bm k}_1) \hat{\zeta}({\bm k}_2) |0 \rangle 
&=& 
- \int_{\tau_{{\rm min},1}}^\tau  d\tau_1 \int_{\tau_{{\rm min},2}}^{\tau_1} d\tau_2\, \langle 0| 
\left[ \left[ \hat{\zeta}^{(0)} ({\bm k}_1, \tau) 
\hat{\zeta}^{(0)} ({\bm k}_2, \tau) , 
H_\zeta (\tau_1 ) \right] , H_\zeta (\tau_2) \right] |0 \rangle  \nonumber\\
&=& \frac{4E_{x}^2}{9\epsilon^2 M_{\rm pl}^4 H^4} \prod_{i=1}^2 
\int_{-1/k_i}^{\tau} \frac{d\tau_i}{\tau_i^4} \left( \tau^3 - \tau_i^3 \right) 
\langle 0|\delta {\cal E}_{x}({\bm k}_1, \tau_1) \delta {{\cal  E}}_{x}
({\bm k}_2, \tau_2) |0 \rangle  \nonumber\\
&=&  \frac{2\pi^2}{k_1^3} \delta^{(3)} ({\bm k}_1 + {\bm k}_2 )  
\frac{E_{x}^2N_k^2 \sin^2 \theta_{{\bm k}_1, {\bm x}}}{\pi^2 \epsilon^2 M_{\rm pl}^4}\,,
\label{powerspe}
\end{eqnarray}
where we used the relation (\ref{commutator1}).
In the first line of Eq.~(\ref{powerspe}) the two integrals have been evaluated 
in the super-horizon regime characterized by $-k_i \tau<1$, 
that is,  $\tau_{{\rm min},i}=-1/k_i$ with $i=1,2$. 
The choice of this contour is based upon the standard vacuum in 
the interacting field theory, that is, the change $\tau \to \tau -i\varepsilon|\tau|$ 
for large $|\tau|$ in the exponent $e^{-ik\tau}$ 
in the mode function (\ref{Alam}) \cite{Maldacena:2002vr}.
This means that the oscillating term in the sub-horizon regime  
is exponentially suppressed, so that the main contribution 
to the integral (\ref{powerspe}) comes from the super-horizon mode 
($-k_i \tau<1$) \cite{Bartolo}. 
In fact, the direct computation of the oscillating contributions to 
the integrals appearing in the correlation functions shows 
that the prescription mentioned above leads to the similar results 
to those derived by the regularization (time averaging) of the oscillating 
terms (see Appendix B of Ref.~\cite{Horn}).
In the second line of Eq.~(\ref{powerspe}) the upper bound $\tau_1$ of 
the second integral has been replaced by $\tau$ by dividing the 
factor 2! because of the symmetry of the integrand. 
We also used the property $\int_{-1/k_i}^{\tau}d\tau_i\,(\tau^3-\tau_i^3)/\tau_i^4
\simeq \ln (aH/k_i) \simeq N_{k_i}$ in the regime $-k_i \tau \ll 1$, where
$N_{k_i}$ is the number of e-foldings before the end of 
inflation at which the modes with the wave number $k_i$ 
left the Hubble radius. 
Since ${\bm k}_1=-{\bm k}_2$, it follows that 
$N_{k_1}=N_{k_2} \equiv N_k$.

Thus, the total scalar power spectrum in the vector model 
is given by  \cite{Bartolo}
\begin{equation}
{\cal P}_{\zeta} = {\cal P}_{\zeta}^{(0)} 
\left(1 + 24  \frac{E_x^2}{\epsilon V} N_k^2  \sin^2 \theta_{{\bm k}_1,{\bm x}}
 \right) = {\cal P}_{\zeta}^{(0)} 
\left( 1  + 48 r_A N_k^2  \sin^2 \theta_{{\bm k}_1,{\bm x}} \right) \,,
\label{Ps_vec}
\end{equation}
where, in the second equality,  we used the relation 
$\rho_A=E_x^2/2$ and the definition $r_A$ given in Eq.~(\ref{rA}).
For the parametrization (\ref{anispe}), 
this result corresponds to a negative anisotropic parameter 
\begin{equation}
g_*=-48\,r_A N_k^2 \,. 
\label{g_vec}
\end{equation}

In order to avoid that the anisotropic contribution does not exceed 
the isotropic spectrum, we demand the condition $|g_*| \lesssim 1$. 
{}From the WMAP data there is the bound
$g_*=0.29 \pm 0.031$ \cite{aniobser3}. 
Under the condition $|g_*| \lesssim 1$ it follows that 
$r_A \lesssim 10^{-5}$ for $N_k \sim 60$.
Since $\alpha$ in Eq.~(\ref{rA}) corresponds to the number of e-foldings 
from the onset of inflation, we have $r_A \simeq 1/(8\alpha_0)=$\,constant 
for $\alpha \ll 10^4$. We recall that, for the coupling (\ref{geneco}), 
the quantity $r_A$ is also constant.
Thus, the scalar spectral index reads
\begin{eqnarray}
n_s-1 &=& \frac{d \ln {\cal P}_{\zeta}}{d \ln k}\bigg|_{k=aH} \nonumber \\
&=& -6\epsilon+2\eta_V +\frac{2}{N_k}\frac{g_*
\sin^2\theta_{{\bm k}_1,{\bm x}}}{1-g_* \sin^2\theta_{{\bm k}_1,{\bm x}}}\,,
\label{ns_vec0}
\end{eqnarray}
where $\eta_V \equiv M_{\rm pl}^2 V_{,\phi\phi}/V$. 
The momentum vector ${\bm k}_1$ does not necessarily need to lie 
on the $(x,y)$-plane, but it is generally given by ${\bm k}_1=
k_1(\sin \theta_1 \cos \varphi_1, \sin \theta_1 \sin \varphi_1, \cos \theta_1)$, 
where $0 \le \theta_1 \le \pi$ and $0 \le \varphi_1 \le 2\pi$. 
It then follows that $\cos \theta_{{\bm k}_1,{\bm x}}=\sin \theta_1 \cos \varphi_1$.
The average of $\sin^2\theta_{{\bm k}_1,{\bm x}}$ integrated over 
all the angles of $\theta_1$ and $\varphi_1$ is
\begin{eqnarray}
\langle \sin^2 \theta_{{\bm k}_1,{\bm x}} \rangle=
\frac{\int_0^{\pi} d\theta_1\,\sin \theta_1 \int_{0}^{2\pi}d \varphi_1 
(1-\sin^2 \theta_1 \cos^2 \varphi_1)}
{\int_0^{\pi} d\theta_1\,\sin \theta_1 \int_{0}^{2\pi}d \varphi_1}
=\frac23\,.
\label{sinave}
\end{eqnarray}
Using this property, 
the scalar spectral index (\ref{ns_vec0}) reads
\begin{equation}
n_s-1 =-6\epsilon+2\eta_V+\frac{4}{N_k} 
\frac{g_*}{3-2g_*}\,.
\label{ns_vec}
\end{equation}

The anisotropic corrections to the two-point 
isotropic correlation of tensor perturbations 
$\hat{h}_s$ $(s=+, \times)$ are
\begin{eqnarray}
\delta \langle 0|\hat{h}_+({\bm k}_1) \hat{h}_+({\bm k}_2) |0 \rangle &=&
- \sum_{A,B=h1,h2} \int_{\tau_{{\rm min},1}}^{\tau} d\tau_1 
\int_{\tau_{{\rm min},2}}^{\tau_1} d\tau_2 \,
\langle 0| \left[ \left[ \hat{h}_+^{(0)}({\bm k}_1, \tau ) 
\hat{h}_+^{(0)}({\bm k}_2, \tau) , H_{A} (\tau_1 ) \right] , H_{B} (\tau_2) \right] |0 \rangle  
\nonumber \\
&=&\frac{4E_x^2}{M_{\rm pl}^4}\frac{\delta^{(3)} 
({\bm k}_1 + {\bm k}_2 ) }{k_1^3} N_k^2 \sin^2\theta_{{\bm k}_1,{\bm x}} \,,
\label{hh_vec}
\end{eqnarray}
where all possible combinations of interacting Hamiltonians $H_{h1}$ and $H_{h2}$ 
are taken, and
\begin{eqnarray}
\delta \langle 0|\hat{h}_{\times}({\bm k}_1) \hat{h}_{\times}({\bm k}_2) |0 
\rangle 
&=& -\int_{\tau_{{\rm min},1}}^{\tau} d\tau_1 
\int_{\tau_{{\rm min},2}}^{\tau_1} d\tau_2 \,
\langle 0| \left[ \left[ \hat{h}_{\times}^{(0)}({\bm k}_1, \tau ) 
\hat{h}_{\times}^{(0)}({\bm k}_2, \tau) , H_{h3} (\tau_1 ) \right] , 
H_{h3} (\tau_2) \right] |0 \rangle    \nonumber \\
&=&\frac{4E_x^2}{M_{\rm pl}^4}\frac{\delta^{(3)} ({\bm k}_1 
+ {\bm k}_2 ) }{k_1^3} N_k^2 \sin^2\theta_{{\bm k}_1,{\bm x}} \,.
\label{hh_vec2}
\end{eqnarray}
Hence we obtain the total correction  
\begin{eqnarray}
\delta \langle 0|\hat{h}_{ij}({\bm k}_1) \hat{h}_{ij}({\bm k}_2) |0 \rangle &=& 
\delta \langle 0|\hat{h}_+({\bm k}_1) \hat{h}_+({\bm k}_2) |0 
\rangle e^{(+)}_{ij} (\bm{k}_1)e^{(+)}_{ij} (\bm{k}_2)+
\delta \langle 0|\hat{h}_{\times}({\bm k}_1) \hat{h}_{\times}({\bm k}_2) 
|0 \rangle e^{(\times)}_{ij} (\bm{k}_1)e^{(\times)}_{ij} (\bm{k}_2)  \nonumber \\
&=& \frac{8E_x^2}{M_{\rm pl}^4}\frac{\delta^{(3)} ({\bm k}_1 + {\bm k}_2 ) }
{k_1^3} N_k^2 \sin^2\theta_{{\bm k}_1,{\bm x}} \,.
\end{eqnarray}
Therefore, the total tensor power spectrum is given by 
\begin{equation}
{\cal P}_{h}=16\epsilon {\cal P}_{\zeta}^{(0)} 
\left( 1+12\epsilon r_A N_k^2  \sin^2 \theta_{{\bm k}_1,{\bm x}} \right) \,.
\label{Ph_vec}
\end{equation}
The tensor-to-scalar ratio can be evaluated by using 
the anisotropic parameter (\ref{g_vec}) as
\begin{equation}
r \equiv \frac{{\cal P}_{h}}{{\cal P}_{\zeta}} =16\epsilon
\frac{1-\epsilon g_* \sin^2 \theta_{{\bm k}_1,{\bm x}}/4}
{1-g_* \sin^2 \theta_{{\bm k}_1,{\bm x}}}\,.
\label{r_vecd}
\end{equation}
Taking the same average over angles as (\ref{sinave}), 
it follows that 
\begin{equation}
r \simeq 16\epsilon \frac{6-\epsilon g_*}{6-4g_*}\,.
\label{r_vec}
\end{equation}
{}From Eq.~(\ref{Ph_vec}) the tensor spectral index reads
\begin{equation}
n_t \equiv \frac{d \ln {\cal P}_h}{d \ln k}\bigg|_{k=aH} 
\simeq -2\epsilon\,,
\label{nt_vec}
\end{equation}
where we neglected the anisotropic contributions because
they are second order in slow-roll parameters.

The cross correlation between curvature perturbations and 
the plus mode of gravitational waves is given by 
\begin{eqnarray}
\langle 0|\hat{\zeta}({\bm k}_1) \hat{h}_+({\bm k}_2) |0 \rangle &=& 
-\sum_{A=h1,h2} \int_{\tau_{{\rm min},1}}^{\tau} d\tau_1 
\int_{\tau_{{\rm min},2}}^{\tau_1} d\tau_2 \,
\langle 0| \left[ \left[ \hat{\zeta}^{(0)}({\bm k}_1, \tau ) \hat{h}_+^{(0)}({\bm k}_2, \tau), 
H_{\zeta} (\tau_1 ) \right] , H_{A} (\tau_2) \right] |0 \rangle + \text{perm.}    \nonumber \\
&=& -\frac{4E_x^2}{\sqrt{2}\epsilon M_{\rm pl}^4}\frac{\delta^{(3)} 
({\bm k}_1 + {\bm k}_2 ) }{k_1^3} N_k^2 \sin^2\theta_{{\bm k}_1,{\bm x}} \,,
\label{zeta_h}
\end{eqnarray}
where ``perm.'' represents the terms obtained by the permutations 
of $H_\zeta$ and $H_A$. 
Similarly we have 
\begin{equation}
\langle 0|\hat{\zeta}({\bm k}_1) \hat{h}_{\times}
({\bm k}_2) |0 \rangle=0\,.
\end{equation}
We define the cross power spectrum ${\cal P}_{\zeta h}(k_1)$ by 
$\langle 0| \hat{\zeta}({\bm k}_1) \hat{h_+}({\bm k}_2)|0 \rangle
=(2\pi^2/k_1^3) \delta^{(3)}({\bm k}_1+{\bm k}_2) 
{\cal P}_{\zeta h}(k_1)$.
While there is no cross correlation without anisotropic interactions, 
it remains for the model (\ref{eq:action}) as
\begin{equation}
{\cal P}_{\zeta h} = - 48\sqrt{2} \, {\cal P}_{\zeta}^{(0)} \epsilon\,r_A N_k^2  
\sin^2 \theta_{{\bm k}_1,{\bm x}} 
=\sqrt{2} \, {\cal P}_{\zeta}^{(0)}\epsilon\,g_* \sin^2 \theta_{{\bm k}_1,{\bm x}} \,.
\end{equation}
This gives rise to the non-vanishing TB cross power spectrum of 
CMB anisotropies \cite{Watanabe2}. 

%%%%%%%%%%%%%%%%%%%%%%%%%%%%%%%%%%%%
\subsection{$f(\phi)^2H_{\mu \nu \lambda}H^{\mu \nu \lambda}$ model}
\label{pertwo}
%%%%%%%%%%%%%%%%%%%%%%%%%%%%%%%%%%%%

Let us proceed to the two-form model given by the action (\ref{oriaction}).
Employing the gauge conditions $B_{i0\,,i}=B_{ij\,,i}=0$, 
we have only one degree of freedom for the two-form field.
The mode expansion can be expressed as
\begin{equation}
B_{ij}({\bm x}, \tau) = B_{ij}^{(0)} + \delta B_{ij}
= B_{ij}^{(0)} +\int \frac{d^3k}{(2\pi)^{3/2}} e^{i{\bm k} \cdot {\bm x}} 
\left[\chi(k,\tau) b({\bm k})
+ \chi^{*}(k,\tau) b^{\dagger} ({-\bm k})
\right]\epsilon_{ij}({\bm k}) \,,
\end{equation}
where $b$, $b^\dagger$ are annihilation and creation operators, $B_{ij}^{(0)}$ is the background value 
with only two non-zero components $B_{yz}^{(0)}=-B_{zy}^{(0)}$,  
and $\epsilon_{ij}({\bm k}) = i\epsilon_{ijl} k^l/(\sqrt{2} k)$ 
is the polarization tensor satisfying the following 
relations\footnote{Compared to the previous paper \cite{Ohashi}, we divided
$\epsilon_{ij}({\bm k})$ by $\sqrt{2}$ to obtain the normalization 
$\epsilon_{ij}({\bm k})\epsilon_{ij}^*({\bm k})=1$. 
We also multiplied the complex factor $i$ to ensure 
the relation $\epsilon_{ij}(- {\bm k}) =\epsilon_{ij}^*({\bm k})$.}
\begin{equation}
k_j \epsilon_{ij}({\bm k}) =0 \,, \quad 
\epsilon_{ij}(- {\bm k}) =\epsilon_{ij}^*({\bm k}) \,, 
\quad
\epsilon_{ij}({\bm k}) \epsilon_{ij}^*({\bm k})=1 \,.
\end{equation}
The tensor $\epsilon_{ij}({\bm k})$ looks similar to 
$e_{ij}^{(s)}({\bm k})$ satisfying the relations 
(\ref{pore1})-(\ref{basis_prop2}), but the difference is that the former 
is anti-symmetric while the latter is symmetric\footnote{If we take 
the momentum vector ${\bm k}_1=k_1 (\cos \theta, \sin \theta, 0)$ 
in the $(x,y)$ plane, $\epsilon_{ij}({\bm k}_1)$ can be expressed as 
\begin{eqnarray}
\epsilon_{ij} (\bm{k}_1) 
= \frac{i}{\sqrt{2}}\left( 
\begin{array}{ccc}
 0 & 0 & - \sin\theta \\
 0 & 0 &  \cos\theta \\
 \sin \theta  &  -\cos \theta & 0
\end{array}
\right)\,.
\label{epij}
\end{eqnarray}
}.

For the coupling (\ref{couplingtwo}), the field $u\equiv f\chi/a$ satisfies 
the following equation of motion on the de Sitter background:
\begin{equation}
u''+\left(k^2 -\frac{2}{\tau^2} \right)u=0\,,
\end{equation}
in which case the scale-invariant spectrum follows.
We can deduce the mode functions as 
\begin{eqnarray}
\chi (k, \tau)=  \frac{H a^3}{\sqrt{k^3}} 
\left( 1+ik\tau \right) e^{-ik\tau} \,. 
\label{uso}
\end{eqnarray}
For convenience we introduce the following variables
\begin{equation}
E_{yz}\equiv \frac{f}{a^3}{B_{yz}^{(0)}}' \,, 
\qquad \delta E_{ij} \equiv \frac{f}{a^3}\delta B_{ij}' \,.
\label{Edef2}
\end{equation}
Then the tree-level interacting Lagrangian is given by
\begin{eqnarray}
L_{\rm int} &=& -\frac{a^4}{12}\left(\langle f^2 \rangle +\frac{\partial \langle f^2 
\rangle}{\partial \phi}\delta \phi\right)(\langle H_{\mu\nu\lambda} \rangle+\delta H_{\mu\nu\lambda})
(\langle H^{\mu\nu\lambda} \rangle+\delta H^{\mu\nu\lambda})  \nonumber \\
&\simeq& a^4 E_{yz} \left(2 \delta E_{yz}\zeta -\delta E_{xz} h_{xy}  
- \delta E_{yz} h_{yy} - \delta E_{yz} h_{zz} + \delta E_{xy} h_{xz}\right) \,,
\label{Lint_two}
\end{eqnarray}
where, in the second line, we picked up the second-order 
terms of perturbations. 

Decomposing the perturbation $\delta E_{ij}({\bm x}, \tau)$ 
into the Fourier modes, as
\begin{equation}
\delta E_{ij} ({\bm x}, \tau)=\int \frac{d^3k}{(2\pi)^{3/2}} 
e^{i {\bm k} \cdot {\bm x}} \delta {\cal E}_{ij} ({\bm k}, \tau)\,,
\label{delEi2}
\end{equation}
the solution in the super-Hubble regime ($|k \tau| \ll 1$) 
is given by 
\begin{equation}
\delta {\cal E}_{ij} ({\bm k}, \tau)=
\frac{3H^2}{\sqrt{k^3}} 
\left[ b ({\bm k})+b^{\dagger} (-{\bm k}) 
\right] \epsilon_{ij}({\bm k})\,.
\end{equation}
The interacting Hamiltonians $H_{\rm int}=-\int d^3 x\,L_{\rm int}$ 
can be written as the sum of the contributions from the five 
terms in Eq.~(\ref{Lint_two}). They are given, respectively, by 
\begin{eqnarray}
H_{\zeta}&=&-\frac{2E_{yz}}{H^4 \tau^4} \int d^3k\, 
\delta {\cal E}_{yz}({\bm k}, \tau) \hat{\zeta}^{(0)}(-{\bm k},\tau) \,, \label{Hzeta2}\\
H_{h1}&=&-\frac{E_{yz}}{\sqrt{2}H^4 \tau^4} \int d^3k\, 
\delta {\cal E}_{xz}({\bm k},\tau) \hat{h}_+^{(0)}(-{\bm k},\tau) 
\sin\theta_{{\bm k}, {\bm x}}\cos\theta_{{\bm k}, {\bm x}} \,, \label{Htwo1}\\
H_{h2}&=&\frac{E_{yz}}{\sqrt{2}H^4 \tau^4} \int d^3k\, 
\delta {\cal E}_{yz}({\bm k},\tau) \hat{h}_+^{(0)}(-{\bm k},\tau) 
\cos^2\theta_{{\bm k}, {\bm x}}  \,, \\
H_{h3}&=&-\frac{E_{yz}}{\sqrt{2}H^4 \tau^4} \int d^3k\, 
\delta {\cal E}_{yz}({\bm k},\tau) \hat{h}_+^{(0)}(-{\bm k},\tau)\,, \label{Htwo3}\\
H_{h4}&=&-\frac{i E_{yz}}{\sqrt{2}H^4 \tau^4} \int d^3k\, 
\delta {\cal E}_{xy}({\bm k},\tau) \hat{h}_{\times}^{(0)}(-{\bm k},\tau) 
\sin\theta_{{\bm k}, {\bm x}}  \,.
\label{Hint_vec1d}
\end{eqnarray}
For the wave number $\bm k$ lying on the $(x,y)$ plane
the component $\delta {\cal E}_{xy}({\bm k},\tau)$ is 0, 
see Eq.~(\ref{epij}). Hence the interacting Hamiltonian $H_{h4}$
associated with the tensor mode $ \hat{h}_{\times}^{(0)}$ vanishes.

The anisotropic contribution $\delta{\cal P}_{\zeta}$ to ${\cal P}_{\zeta}^{(0)}$ 
can be evaluated from the interacting Hamiltonian $H_{\zeta}$. 
The total scalar power spectrum has been derived in Ref.~\cite{Ohashi}, as
\begin{equation}
{\cal P}_{\zeta} = {\cal P}_{\zeta}^{(0)} \left( 1  + 6  \frac{E_{yz}^2}
{\epsilon V} N_k^2  \cos^2 \theta_{{\bm k}_1,{\bm x}} \right)
={\cal P}_{\zeta}^{(0)} \left( 1  + 12  r_B N_k^2  
\cos^2 \theta_{{\bm k}_1,{\bm x}} \right) \,,
\label{Ps_two}
\end{equation}
where, in the second equality, we used $\rho_B=E_{yz}^2/2$ 
and the definition $r_B$ given in Eq.~(\ref{rB}).
Comparing Eq.~(\ref{Ps_two}) with Eq.~(\ref{anispe}), 
the anisotropic parameter can be expressed as
\begin{equation}
g_*=12 r_B N_k^2 \,,
\label{g_two}
\end{equation}
which is positive unlike the vector model.
Since $r_B$ is nearly constant for $\alpha \ll \alpha_0$, 
we obtain the scalar spectral index 
\begin{eqnarray}
n_s-1 &=& \frac{d \ln {\cal P}_{\zeta}}{d \ln k}\bigg|_{k=aH} \nonumber \\
&=& -6\epsilon+2\eta_V -\frac{2}{N_k}
\frac{g_*\cos^2\theta_{{\bm k}_1,{\bm x}}}{1+g_* 
\cos^2\theta_{{\bm k}_1,{\bm x}}} \nonumber \\
&=& -6\epsilon+2\eta_V -\frac{2}{N_k}\frac{g_*}{3+g_*} \,.
\label{ns_two}
\end{eqnarray}
In the last equality we used the fact that the average of 
$\cos^2\theta_{{\bm k}_1,{\bm x}}$
integrated over all the angles is 1/3.

Summing up all possible combinations of the interacting Hamiltonians 
$H_{h1}$, $H_{h2}$, and $H_{h3}$, the anisotropic correction from 
the $\hat{h}_+$ mode to the tensor power spectrum 
is\footnote{The perturbations $\delta {\cal E}_{xz} ({\bm k},\tau)$ and 
$\delta {\cal E}_{yz} ({\bm k},\tau)$ appearing in Eqs.~(\ref{Htwo1})-(\ref{Htwo3}) 
have the basis components $-i \sin \theta_{{\bm k},{\bm x}}/\sqrt{2}$ and 
$i \cos \theta_{{\bm k},{\bm x}}/\sqrt{2}$, respectively, see Eq.~(\ref{epij}).	 
It then follows that $H_{h1}+H_{h2}+H_{h3}=0$. 
Hence the sum of interacting Hamiltonians between the two-form 
field and tensor perturbations vanishes.}
\begin{eqnarray}
 \delta \langle 0|\hat{h}_+({\bm k}_1) \hat{h}_+({\bm k}_2) |0 \rangle &=& 
- \sum_{A,B=h1,h2,h3}\int^{\tau}_{\tau_{{\rm min},1}} d\tau_1 \int_{\tau_{{\rm min},2}}^{\tau_1} d\tau_2 \,
 \langle 0| \left[ \left[ \hat{h}_+^{(0)}({\bm k}_1, \tau ) \hat{h}_+^{(0)}({\bm k}_2, \tau) , H_{A} (\tau_1 ) \right] , H_{B} (\tau_2) \right] |0 \rangle    \nonumber \\
 &=&\frac{4E_{yz}^2}{M_{\rm pl}^4}\frac{\delta^{(3)} ({\bm k}_1 + {\bm k}_2 ) }{k_1^3} N_k^2 
 \left(\sin^4\theta\cos^2\theta+\cos^6\theta+\cos^2\theta+2\sin^2\theta\cos^4\theta-2\sin^2\theta\cos^2\theta-2\cos^4\theta\right)  \nonumber \\
  &=& 0 \,.
\label{hh_two}
\end{eqnarray}
We also have 
$\delta \langle 0|\hat{h}_{\times}({\bm k}_1) \hat{h}_{\times}({\bm k}_2) |0 \rangle=0$
because of the property $H_{h4}=0$.
Since $\delta {\cal P}_h=0$, it follows that ${\cal P}_h={\cal P}_h^{(0)}$.
Thus, the tensor-to-scalar ratio is given by 
\begin{equation}
r=16\epsilon \frac{1}{1+g_* \cos^2 \theta_{{\bm k}_1,{\bm x}}}
=16\epsilon \frac{3}{3+g_*}\,,
\label{r_two}
\end{equation}
where, in the last equality, we have taken the average over all the angles. 
The tensor spectral index reads
\begin{equation}
n_t=-2\epsilon\,.
\label{nt_two}
\end{equation}

Similarly, the cross correlations between scalar 
and tensor perturbations are computed to give
\begin{equation}
\langle 0|\hat{\zeta}({\bm k}_1) \hat{h}_+({\bm k}_2) |0 \rangle =0  \,, 
\qquad 
\langle 0|\hat{\zeta}({\bm k}_1) \hat{h}_{\times}({\bm k}_2) |0 \rangle=0\,,
\end{equation}
which mean that the cross power spectrum ${\cal P}_{\zeta h}$ is 0. 
This is an interesting property by which the two-form model 
can be distinguished from the vector model.

%===============================================================%
%************************ SECTION IV ****************************%
%===============================================================%

%%%%%%%%%%%%%%%%%%%%%%%%%%%%%%%%%%%%%
\section{Joint observational constraints on anisotropic inflationary models}
\label{obs_vec}
\label{obssec}
%%%%%%%%%%%%%%%%%%%%%%%%%%%%%%%%%%%%%

In this section, we place observational constraints on each anisotropic 
inflationary model with concrete inflaton potentials. 
Using the Cosmological Monte Carlo (CosmoMC) code \cite{cosmomc,Lewis}, 
we carry out the likelihood analysis with the latest Planck data \cite{Adecosmo} 
combined with the WMAP large-angle polarization (WP) \cite{WMAP9}, 
Baryon Acoustic Oscillations (BAO) \cite{BAO1,BAO2,BAO3}, 
and ACT/SPT temperature data of high multipoles (high-$\ell$) \cite{Das,Rei}.
The flat $\Lambda$CDM model is assumed with $N_{\rm eff} = 3.046$
relativistic degrees of freedom and with the instant reionization. 
We also set the runnings of the scalar and tensor spectral indices to be 0.
The pivot wave number is chosen to be $k_0=0.05$~Mpc$^{-1}$.
We confirmed that the different choices of $k_0$ such as $0.002$~Mpc$^{-1}$ 
give practically identical likelihood results.

{}From Eqs.~(\ref{r_vec}), (\ref{nt_vec}), and (\ref{r_two}), (\ref{nt_two}), 
the consistency relations in the two anisotropic models are given by 
\begin{eqnarray}
r &=& -8n_t \frac{6-\epsilon g_*}{6-4g_*} \qquad (\text{vector~model}),\\
r &=& -8n_t \frac{3}{3+g_*} \qquad~\,(\text{two-form model}).
\end{eqnarray}
The presence of anisotropic interactions modifies the standard 
consistency relation $r=-8n_t$. 
If $g_*=-0.5$ and $g_*=0.5$, then we have 
$r \simeq -6.0n_t$ for the vector model and $r \simeq -6.9n_t$ 
for the two-form model, respectively. 
We have run the CosmoMC code by using these consistency relations
and found that the likelihood contours are very similar to those derived
with the relation $r=-8n_t$. 
Therefore, we plot observational contours obtained by varying the 
three inflationary observables ${\cal P}_{\zeta} (k_0)$, $n_s(k_0)$, 
and $r(k_0)$ with the consistency relation $r(k_0)=-8n_t(k_0)$. 

In the vector model, anisotropic interactions lead to the enhancement of
the scalar power spectrum ${\cal P}_{\zeta}$ on larger scales 
because the amplitude (\ref{Ps_vec}) increases for larger $N_k$.
As a result, the spectral index $n_s$ gets smaller for any inflaton potentials.
The power spectrum of gravitational waves is also enhanced in the 
presence of anisotropic sources, but its effect is small compared to 
that on ${\cal P}_{\zeta}$. Hence the tensor-to-scalar ratio gets smaller
irrespective of the inflaton potentials.
We recall that the decreases of $n_s$ and $r$ are controlled by 
the negative anisotropic parameter $g_*$ given in Eq.~(\ref{g_vec}).

In the two-form model, anisotropic interactions also lead to the decrease 
of $n_s$ and $r$ with the positive parameter $g_*$ given in 
Eq.~(\ref{g_two}).
Since the level of the enhancement of ${\cal P}_{\zeta}$ 
is different from that of the vector model,
the observables $n_s$ and $r$ exhibit some difference 
between the two anisotropic inflation models.

In the following we study the vector and two-form models 
separately for several different inflaton potentials. 
Since we are considering the case $c=1$, we can employ 
the standard slow-roll equations (\ref{Friap}) and 
$3H \dot{\phi} \simeq -V_{,\phi}$ at the background level. 
Under this approximation, the number of e-foldings is given by
\begin{equation}
N_k \simeq \frac{1}{M_{\rm pl}^2} 
\int_{\phi_f}^{\phi} \frac{V}{V_{,\tilde{\phi}}}d \tilde{\phi}\,,
\label{efold}
\end{equation}
where $\phi_f$ is the value of $\phi$ at the end of inflation 
determined by the condition $\epsilon (\phi_f)=1$. 
For the comparison of the inflationary observables with 
the CMB data, we fix $N_k=60$.

%%%%%%%%%%%%%%%%%%%%%%%%%%%
\subsection{$f(\phi)^2F_{\mu \nu}F^{\mu \nu}$ model}
%%%%%%%%%%%%%%%%%%%%%%%%%%%

Let us study observational constraints on the vector model.

First, we consider chaotic inflation characterized 
by the power-law potential \cite{chaotic}
\begin{equation}
V(\phi)=\lambda\phi ^n/n \,,
\label{chao_pot}
\end{equation}
where $n$ and $\lambda$ are positive constants.
In this case, we have that $\epsilon=n^2M_{\rm pl}^2/(2\phi ^2)$ and 
$\eta_V=n(n-1)M_{\rm pl}^2/\phi^2$.
The field value at the end of inflation can be estimated as 
$\phi_f=nM_{\rm pl}/\sqrt{2}$. From Eq.~(\ref{efold}) the number of 
e-foldings $N_k$ is related to the field $\phi$, 
as $\phi^2\simeq2n(N_k+n/4)M_{\rm  pl}^2$.
Then the observables (\ref{ns_vec}) and (\ref{r_vec}), which are 
averaged over all the angles, reduce to 
\begin{equation}
n_s=1-\frac{6N_k(n+2)-4 \left[ N_k(n+6)+n \right] g_*}
{N_k(n+4N_k)(3-2g_*)}\,, 
\qquad r= \frac{8n[6(n+4N_k)-ng_*]}{(n+4N_k)^2(3-2g_*)}\,.
\label{chaoticob}
\end{equation}

\begin{figure}
\includegraphics[height=3.3in,width=3.5in]{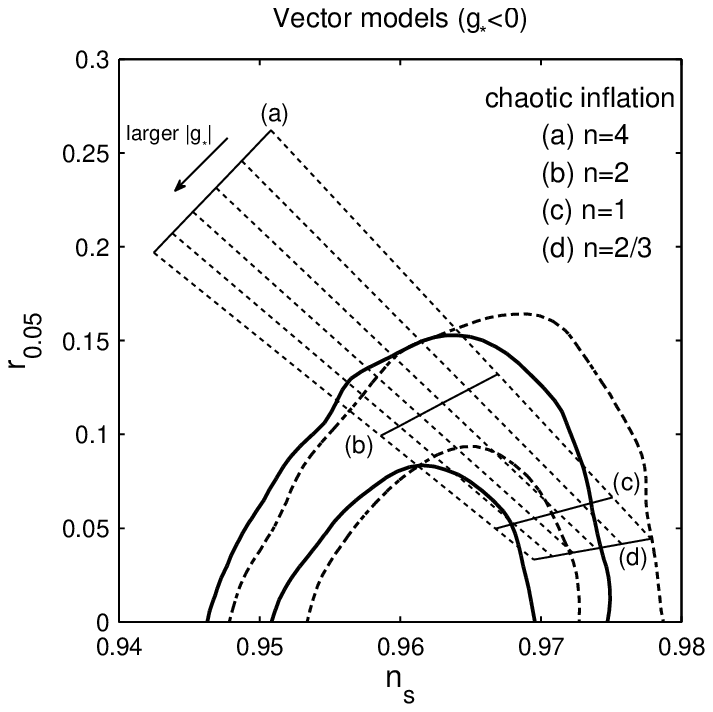}
\includegraphics[height=3.3in,width=3.5in]{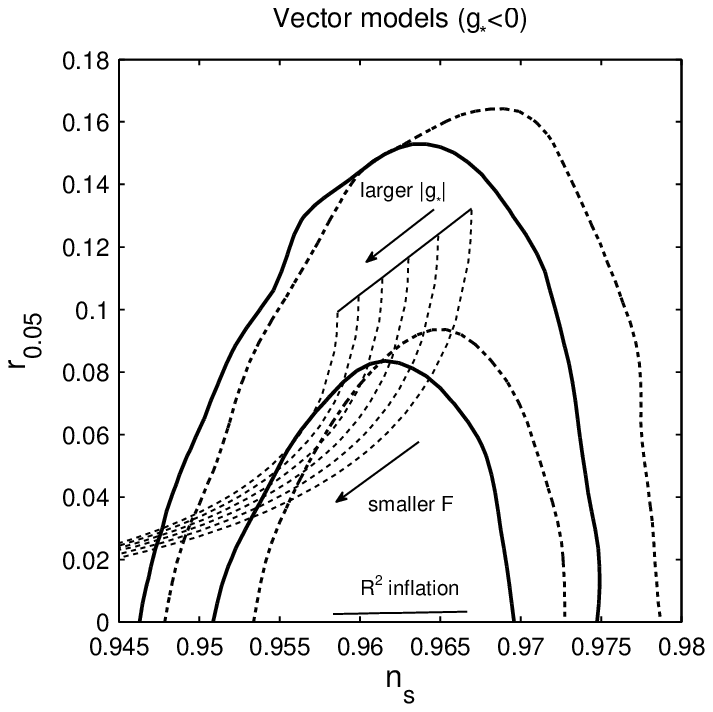}
\caption{\label{fig1}
2-dimensional observational bounds in the $(n_s, r)$ plane with the number of e-foldings
$N_k = 60$ and the pivot wave number  $k_0=0.05$~Mpc$^{-1}$.
The bold solid curves represent the 68\,\%\,CL (inside) and 
95\,\%\,CL (outside) boundaries derived by 
the joint data analysis of the Planck+WP+BAO+high-$\ell$ data, 
whereas the thick dashed curves correspond to the contours constrained 
by the Planck+WP+BAO data. 
In both cases the consistency relation $r(k_0)=-8n_t(k_0)$ is used.
We consider several different inflaton potentials in the vector model, i.e.,  
(i) $V(\phi)=\lambda \phi^n/n$ ($n=4,2,1,2/3$) (left), 
(ii) $V(\phi)=V_0 [1+\cos(\phi/F)]$ (right), and 
(iii) $V(\phi)=(3/4)M^2M_{\rm pl}^2 ( 1-e^{-\sqrt{2/3} \phi/M_{\rm pl}} )^2$ (right), 
with $g_*$ ranging $-0.5 \leq g_* \leq 0$. 
The thin dotted curves correspond to the anisotropic parameters 
$g_*=0,-0.1,-0.2,-0.3,-0.4,-0.5$, respectively. 
The thin line labelled as ``$R^2$ inflation'' shows the theoretical plots 
predicted by the potential (\ref{pot_S}).
In the presence of the vector field, both $n_s$ and $r$ get smaller.}
\end{figure}

In the left panel of Fig.~\ref{fig1} we plot the theoretical values of $n_s$ and $r$ 
for the anisotropic parameter $g_*$ ranging in the region $-0.5 \leq g_* \leq 0$ 
with $N_k=60$. The self-coupling potential $V(\phi)=\lambda \phi^4/4$ 
is outside the 95\,\%\,confidence level (CL) observational boundaries even 
in the presence of anisotropic interactions. 
The quadratic potential $V(\phi)=\lambda \phi^2/2$ is inside the 95\,\%\,CL boundaries, 
but it is still outside the 68\,\%\,CL contours.
When $g_*=0$, the linear potential $V(\phi)=\lambda \phi$, which appears in the axion 
monodromy scenario \cite{mono1}, is outside the 95\,\%\,CL boundary constrained 
by the Planck+WP+BAO+high-$\ell$ data, but the vector anisotropy with 
$g_*<-0.4$ allows the model to be inside the 68 \,\%\,CL contour. 
A similar property also holds for another axion monodromy potential 
$V(\phi)=(3\lambda/2)\phi^{2/3}$ \cite{mono2}, 
but larger values of $|g_*|$ are required for the compatibility with the data.

%%%%%%%%%%%%%%%%%%%%%%%%%%%%%%%%%%%
%%%%%%%%%%%%%%%%%%%%%%%%%%%%%%%%%

We also study natural inflation characterized by the potential \cite{Freese}
\begin{equation}
V(\phi)=\Lambda^4 \left[ 1+\cos(\phi/F) \right] \,,
\label{nat_pot}
\end{equation}
where $\Lambda$ and $F$ are constants having a dimension of mass.
The relation between $N_k$ and $\phi$ is given by
$N_k=(2F^2/M_{\rm pl}^2)\ln\left[ \sin(\phi_f/(2F)) / \sin(\phi/(2F)) \right]$,
where $\phi_f$ is known by solving the equation 
$\tan^2\left[ \phi_f/(2F) \right] = 2(F/M_{\rm pl})^2$.
The observables (\ref{ns_vec}) and (\ref{r_vec}) read
\begin{eqnarray}
n_s&=&1-\frac{3N_k M_{\rm pl}^2[3-\cos(\phi/F)]
-\{2N_k M_{\rm pl}^2[3-\cos(\phi/F)]+4F^2[1+\cos(\phi/F)] \} g_*}
{[1+\cos(\phi/F)]F^2N_k(3-2g_*)} \,, \label{ns_nat}\\
r&=&\frac{2\{12F^2[1+\cos(\phi/F)]-M_{\rm pl}^2[1-\cos(\phi/F)]g_*\}
[1-\cos(\phi/F)]M_{\rm pl}^2}{[1+\cos(\phi/F)]^2F^4(3-2g_*)} \,. 
\label{r_nat}
\end{eqnarray}
For a given value of $F$ we can numerically identify the field value $\phi$ 
corresponding to $N_k=60$. Then, we evaluate $n_s$ and $r$ according 
to the formulas (\ref{ns_nat}) and (\ref{r_nat}).
In the limit that $F\rightarrow \infty$, these observables approach the values 
(\ref{chaoticob}) of chaotic inflation with $n = 2$.
In the right panel of Fig.~\ref{fig1}, we show the theoretical values of $n_s$ 
and $r$ for different values of $F$ and $g_*$.
For smaller $F$ and larger $|g_*|$, both $n_s$ and $r$ get smaller.
When $g_*=0$, the mass scale $F$ is constrained to be 
$5.1M_{\rm pl}<F<7.9M_{\rm pl}$ (68\,\%\,CL) from the 
Planck+WP+BAO+high-$\ell$ data \cite{Kuro}. 
For larger $|g_*|$, the allowed parameter space inside the 68\,\%\,CL 
contours tends to be narrower.
In particular, if $|g_*|>0.5$, then the model is outside the 68\,\%\,CL 
boundary constrained by the Planck+WP+BAO+high-$\ell$ data.
Thus, in natural inflation, the presence of anisotropic interactions
leads to the deviation from the observationally favored region.

Let us also discuss the inflaton potential of the form
\begin{equation}
V(\phi)=\frac{3}{4}M^2M_{\rm pl}^2 
\left( 1-e^{-\sqrt{2/3} \phi/M_{\rm pl}} \right)^2\,,
\label{pot_S}
\end{equation}
where $M$ is a constant having a dimension of mass. 
This potential arises in the Starobinsky's model 
$f(R)=R+R^2/(6M^2)$ \cite{Starobinsky}
after a conformal transformation to the Einstein frame with 
the field definition $\phi/M_{\rm pl}=\sqrt{3/2}\ln[\partial f(R)/\partial R]$ \cite{Maeda}.
Recently there have been numerous attempts to construct the potential 
(\ref{pot_S}) in the context of supergravity and quantum 
gravity \cite{Ketov}.
In the regime $\phi/M_{\rm pl}\gg 1$, the number of e-foldings is 
related to the inflaton, as $e^{-\sqrt{2/3} \phi/M_{\rm pl}} \simeq 3/(4N_k)$ \cite{DeFelice}.
The slow-roll parameters are approximately given by 
$\epsilon\simeq 3/(4N_k^2)$ and $\eta_V\simeq -1/N_k$, 
which means that $\epsilon$ is much smaller than $|\eta_V|$.
Therefore, the observables (\ref{ns_vec}) and (\ref{r_vec}) reduce to 
\begin{equation}
n_s=1-\frac{2(3-4g_*)}{N_k(3-2g_*)} \,, 
\qquad r=\frac{9(8N_k^2-g_*)}{2N_k^4(3-2g_*)} \,.
\end{equation}
When $g_*=0$ we have $n_s=1-2/N_k$ and $r=12/N_k^2$, which 
correspond to the values in the Starobinsky's model \cite{Staper}. 
The anisotropic interactions lead to the decrease of $n_s$, 
but still the model is well inside the 68\,\%\,CL contour, see 
the right panel of Fig.~\ref{fig1}.

We also study hybrid inflation characterized by the potential 
$V(\phi)=\Lambda^4+m^2\phi^2/2$ \cite{hybrid}, 
where $\Lambda$ and $m$ are constants.
When $g_*=0$, this model gives rise to a blue-tilted spectrum ($n_s>1$).
In the presence of anisotropic interactions 
it is possible to have a red-tilted spectrum, 
but we find that $n_s$ is larger than 0.99 for $|g_{*}|<0.5$.
Hence the model is still outside the 95\,\%\,CL region.

%%%%%%%%%%%%%%%%%%%%%%%%%%%%%%%%%%%%
\subsection{$f(\phi)^2H_{\mu \nu \lambda}H^{\mu \nu \lambda}$ model}
%%%%%%%%%%%%%%%%%%%%%%%%%%%%%%%%%%%%

%%%%%%%%%%%%%%%%%%%%%%%%%%%%%%%%%%%
\begin{figure}
\includegraphics[height=3.3in,width=3.5in]{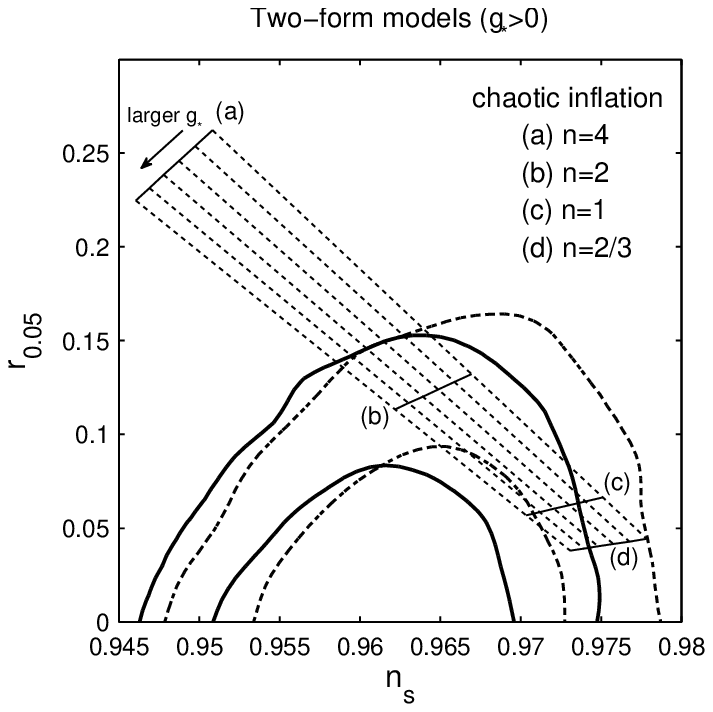}
\includegraphics[height=3.3in,width=3.5in]{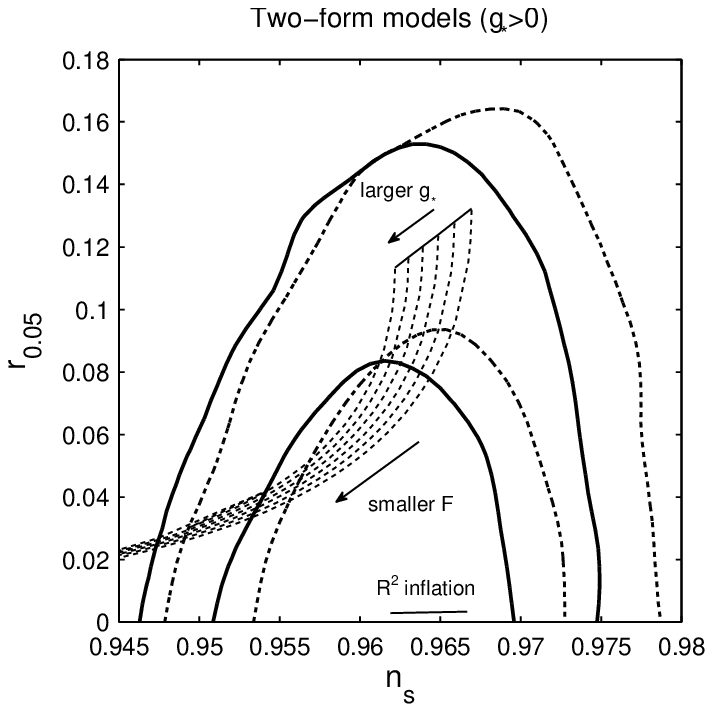}
\caption{\label{fig2}
The same observational contours as those given in Fig.~\ref{fig1} 
(denoted as thick solid/dashed curves). 
The thin solid and dotted lines show the theoretical predictions 
of the two-form model with $N_k=60$ for three 
inflaton potentials (\ref{chao_pot}), (\ref{nat_pot}), and (\ref{pot_S}).
We choose five different values of the anisotropic parameter: 
$g_*=0, 0.1, 0.2, 0.3, 0.4, 0.5$.
For larger $g_*$, both $n_s$ and $r$ get smaller.}
\end{figure}
%%%%%%%%%%%%%%%%%%%%%%%%%%%%%%%%%

In the two-form model the scalar spectral index and the tensor-to-scalar
ratio, which are averaged over all the angles, are given by Eqs.~(\ref{ns_two}) 
and (\ref{r_two}), respectively.

For chaotic inflation characterized by the potential 
(\ref{chao_pot}), these observables reduce to 
\begin{equation}
n_s=1-\frac{6N_k(n+2)+2[N_k(n+6)+n] g_*}
{N_k(n+4N_k)(3+g_*)}\,, 
\qquad 
r=  \frac{48n}{(n+4N_k)(3+g_*)}\,.
\end{equation}
In the left panel of Fig.~\ref{fig2} the theoretical predictions of 
chaotic inflation are shown for $g_*$ ranging 
in the region $0 \leq g_* \leq 0.5$. 
For larger $g_*$ both $n_s$ and $r$ decrease, but the quadratic 
potential is outside the 68\,\%\,CL region. 
In the presence of anisotropic interactions the potentials 
with $n=1$ and $n=2/3$ enter the 95\,\%\,CL boundaries, 
but still these models are outside the 68\,\%\,CL region 
constrained by the Planck+WP+BAO+high-$\ell$ data.
This shows that, for the same value of $|g_*|$, the power-law 
potentials with $n \leq 1$ in the two-form model are more 
difficult to be compatible with the data relative to the same
potentials in the vector model.

For natural inflation given by the potential (\ref{nat_pot}),
the observables (\ref{ns_two}) and (\ref{r_two}) read  
\begin{eqnarray}
n_s &=& 
1-\frac{3N_k M_{\rm pl}^2 [3- \cos(\phi/F)]+
\{ N_k M_{\rm pl}^2[3- \cos(\phi/F) ]+2F^2[1+\cos(\phi/F)] \}g_* }
{[ 1+\cos(\phi/F)]F^2N_k (3+g_*)} \,, \\
r &=& \frac{24M_{\rm pl}^2 [1- \cos(\phi/F)]}
{F^2(3+g_*)[1+ \cos(\phi/F)]}\,,
\end{eqnarray}
which decrease for larger $g_*$ and smaller $F$. 
The difference from the vector model is that, for 
the same value of $|g_*|$, the allowed region of the 
two-form model inside the 68\,\%\,CL observational 
contours is wider.
When $g_*=0.5$, for example, we find that the mass scale $F$
is constrained to be $5.9M_{\rm pl}<F<10.1M_{\rm pl}$ (68\,\%\,CL) 
in the two-form model.
As long as $g_*$ is smaller than $1.0$, there are some allowed 
values of $F$ compatible with the observational data at 68\,\%\,CL.

For the potential (\ref{pot_S}) we have
\begin{equation}
n_s=1-\frac{2(3+2g_*)}{N_k(3+g_*)} \,, \qquad r=\frac{36}{N_k^2(3+g_*)} \,.
\end{equation}
As we see in the right panel of Fig.~\ref{fig2}, 
this potential is well within the 68\,\%\,CL region.

We also find that hybrid inflation with the 
potential $V(\phi)=\Lambda^4+m^2\phi^2/2$ is
outside the 95\,\%\,CL boundaries for $g_*<0.5$.

In summary, the observables $n_s$ and $r$ in the two-form model 
exhibit similar properties as those discussed in the vector model, 
but there are some potentials which can be inside and outside 
the 68\,\%\,CL region depending 
on the values of $g_*$. 
The precise observational bounds of $g_*$ (in particular the signs of $g_*$) 
will be able to distinguish the two anisotropic
inflationary models further.

%===============================================================%
%************************ SECTION V ****************************%
%===============================================================%

%%%%%%%%%%%%%%%%%%%%%%%%%
\section{Statistically anisotropic non-Gaussianities}
\label{nonsec}
%%%%%%%%%%%%%%%%%%%%%%%%%

In this section, we estimate anisotropic contributions to the statistical 
non-Gaussianities of curvature perturbations.
The three-point correlation functions of $\zeta$ have been already derived 
both in the vector model \cite{Bartolo} and in the two-form model \cite{Ohashi}.
In the vector model, Bartolo {\it et al.} \cite{Bartolo} considered the squeezed shape 
in which the angles $\theta_{{\bm k}_1 ,{\bm k}_3}$ and 
$\theta_{{\bm k}_2 ,{\bm k}_3}$ between the momentum vectors satisfying 
the relation ${\bm k}_1+{\bm k}_2+{\bm k}_3=0$ are not necessarily close to
$\pi/2$ and they took the average over angles to evaluate the local non-linear 
estimator $f_{\rm NL}^{\rm local}$.
In the two-form model, the present authors \cite{Ohashi} took the 
strict squeezed limit, $k_3 \to 0$, $\theta_{{\bm k}_1 ,{\bm k}_3} \to \pi/2$,  
$\theta_{{\bm k}_2 ,{\bm k}_3} \to \pi/2$, and showed that $f_{\rm NL}^{\rm local}$ 
vanishes. We clarify the difference between these two approaches and estimate 
the local non-linear estimators in both analyses.

In the two-form model, the non-linear estimators have been also computed for 
other shapes of non-Gaussianities such as the equilateral 
and enfolded ones \cite{Ohashi}.
Since the similar analysis has not been yet done in the vector model,  
we evaluate $f_{\rm NL}$ in the equilateral 
and enfolded limits.

%%%%%%%%%%%%%%%%%%%%%%%%%%%
\subsection{$f(\phi)^2F_{\mu \nu}F^{\mu \nu}$ model}
%%%%%%%%%%%%%%%%%%%%%%%%%%%

In order to calculate the three-point correlation function of curvature perturbations 
in the vector model described by the action (\ref{eq:action}), 
we need to derive the Hamiltonian $H_{\zeta 2}$ following from the third-order 
interacting Lagrangian $L_{\rm int}^{(3)}\simeq 2a^4\delta E_i \delta E_j \zeta$, 
in addition to $H_{\zeta}$ given in Eq.~(\ref{Hzeta}). 
On using Eqs.~(\ref{zeta0def}) and (\ref{delEi}), 
this interacting Hamiltonian is given by
\begin{equation}
H_{\zeta2}=-\frac{2}{H^4\tau^4}\int \, \frac{d^3k d^3p}{(2\pi)^{3/2}} \, 
\delta{\cal E}_i({\bm k},\tau) \delta {\cal E}_j({\bm p},\tau) 
\hat{\zeta}^{(0)}(-{\bm k}-{\bm p},\tau) \,.
\label{Hzeta3}
\end{equation}

Then we can evaluate the three-point correlation of $\zeta$, as \cite{Bartolo}
\begin{eqnarray}
\delta \langle 0|\hat{\zeta}({\bm k}_1) \hat{\zeta}({\bm k}_2)\hat{\zeta}({\bm k}_3) |0 \rangle 
&=& 
i \int_{\tau_{{\rm min},1}}^\tau  d\tau_1 \int_{\tau_{{\rm min},2}}^{\tau_1} 
d\tau_2\, \int_{\tau_{{\rm min},3}}^{\tau_2} d\tau_3\, \nonumber \\
&& \times \langle 0| 
\left[\left[ \left[ \hat{\zeta}^{(0)} ({\bm k}_1) 
\hat{\zeta}^{(0)} ({\bm k}_2) \hat{\zeta}^{(0)} ({\bm k}_3)(\tau) , 
H_{\zeta2} (\tau_1 ) \right] , H_\zeta (\tau_2) \right] H_\zeta (\tau_3)  
\right] |0 \rangle 
+ \text{2 perm.}  \nonumber\\
&=& \frac{4E_{x}^2}{27\epsilon^3 M_{\rm pl}^6 H^6} \prod_{i=1}^3 
\int_{-1/k_i}^{\tau} \frac{d\tau_i}{\tau_i^4} \left( \tau^3 - \tau_i^3 \right) 
\nonumber \\ 
&& \times \int \, \frac{d^3p}{(2\pi)^{3/2}}
\langle 0|\delta {\cal E}_{x}({\bm k}_1, \tau_1) \delta {{\cal  E}}_{x}({\bm k}_2, \tau_2) 
\delta {\cal E}_{i}({\bm p}, \tau_3) \delta {{\cal  E}}_{j}({\bm k}_3-{\bm p}, \tau_3) |0 \rangle 
+ \text{2 perm.}  \nonumber\\
&\simeq&  288\sqrt{2}\pi^{5/2} \frac{E_{x}^2}{\epsilon V} N_{k_1} N_{k_2} N_{k_3} ({\cal P}_{\zeta}^{(0)})^2 \delta^{(3)} ({\bm k}_1 + {\bm k}_2 + {\bm k}_3) \nonumber \\
&&\times \left[\frac{1}{k_1^3 k_2^3}( 1- \cos^2 \theta_{{\bm k}_1, {\bm x}} -\cos^2\theta_{{\bm k}_2, {\bm x}}+ \cos\theta_{{\bm k}_1,  {\bm x}}\cos\theta_{{\bm k}_2,  {\bm x}}\cos\theta_{{\bm k}_1, {\bm k}_2})
+{\rm 2~perm.}\right].
\label{bispe}
\end{eqnarray}
The total anisotropic bispectrum ${\cal B}_{\zeta}$ is defined 
by $\delta \langle 0|\hat{\zeta}({\bm k}_1) \hat{\zeta}({\bm k}_2)
\hat{\zeta}({\bm k}_3) |0 \rangle={\cal B}_{\zeta}\delta^{(3)} ({\bm k}_1 + {\bm k}_2 + {\bm k}_3) $.
We introduce the non-linear estimator $f_{\rm NL}$ in the form
\begin{equation}
{\cal B}_{\zeta}=\frac{3}{10}(2\pi)^{5/2}f_{\rm NL} 
({\cal P}_{\zeta})^2
\sum_{i=1}^3 k_i^3 / \prod_{i=1}^3 k_i^3 \,,
\label{def_fnl}
\end{equation}
by which $f_{\rm NL}$ for the vector model can be derived as \cite{Bartolo}
\begin{eqnarray}
\hspace{-0.7cm}
f_{\rm NL}&=&480\,r_A \frac{({\cal P}_{\zeta}^{(0)})^2}
{({\cal P}_{\zeta})^2} \frac{N_{k_1} N_{k_2} N_{k_3}}{k_1^3+k_2^3+k_3^3}
[k_3^3( 1- \cos^2 \theta_{{\bm k}_1, {\bm x}} -\cos^2\theta_{{\bm k}_2, {\bm x}}+ \cos\theta_{{\bm k}_1,  {\bm x}}\cos\theta_{{\bm k}_2,  {\bm x}}\cos\theta_{{\bm k}_1, {\bm k}_2})+{\rm 2~perm.}] \nonumber \\
\hspace{-0.7cm}
&\simeq& 60\left(\frac{-g_*}{0.1}\right)\left(\frac{N_k}{60}\right)\frac{1}{k_1^3+k_2^3+k_3^3} 
[k_3^3( 1- \cos^2 \theta_{{\bm k}_1, {\bm x}} -\cos^2\theta_{{\bm k}_2, {\bm x}}+ \cos\theta_{{\bm k}_1,  {\bm x}}\cos\theta_{{\bm k}_2,  {\bm x}}\cos\theta_{{\bm k}_1, {\bm k}_2})+{\rm 2~perm.}],
\label{fnl_vec}
\end{eqnarray}
where, in the first line, we used $\rho_A=E_x^2/2$ and the 
quantity $r_A$ defined in Eq.~(\ref{rA}).
In the second line we employed the approximations 
$({\cal P}_{\zeta})^2\simeq ({\cal P}_{\zeta}^{(0)})^2$,  
$N_{k_1}\simeq N_{k_2}\simeq N_{k_3} \equiv N_k$, and
the definition of $g_*$ given in Eq.~(\ref{g_vec}).

Let us first take the same approach as that used 
in Ref.~\cite{Ohashi} for the two-form model.
It is convenient to introduce the following parameters
\begin{equation}
r_2\equiv\frac{k_2}{k_1}\,, \quad r_3\equiv\frac{k_3}{k_1} \,.
\label{r2_r3}
\end{equation}
If we fix $r_2 = 1$ and define the angles 
$\beta=\pi-\theta_{{\bm k}_1, {\bm k}_2}$ and $\gamma=\theta_{{\bm k}_1, {\bm x}}$, 
it follows that 
\begin{eqnarray}
\hspace{-0.2cm}
f_{\rm NL} 
&\simeq& 60 \left(\frac{-g_*}{0.1}\right)\left(\frac{N_k}{60}\right)\frac{1}{2+r_3^3} 
\biggl[ \frac{1}{2}-r_3^3 \cos\beta (\sin\beta \sin\gamma \cos\gamma-\cos\beta+\cos\beta \cos^2\gamma)
 -\cos^2\beta\cos^2\gamma \nonumber \\
\hspace{-0.2cm}&&~~~~~~~~~~~~~~~~~~~~~~~~~~~~~~~~~
+\frac{1}{2}\cos^2\beta+\cos\beta\cos^2\gamma-\cos\beta\sin\beta\cos\gamma\sin\gamma
-\cos\beta+\sin\beta\sin\gamma\cos\gamma \biggr].
\label{fnl_vec2}
\end{eqnarray}
The angle $\beta$ is in the range $0 < \beta < \pi$ (i.e., $0 < r_3 < 2$).
In the strict squeezed limit ($r_3 \to 0$ and $\beta \to 0$), the local estimator 
$f_{\rm NL}^{\rm local}$ vanishes for any values of $\gamma$.
This limit corresponds to the case in which the angles 
$\theta_{{\bm k}_2, {\bm k}_3}$ and $\theta_{{\bm k}_3, {\bm k}_1}$ approach $\pi/2$.

Bartolo {\it et al.} \cite{Bartolo} estimated $f_{\rm NL}^{\rm local}$ in the following way.
For the incomplete squeezed shape the angles $\theta_{{\bm k}_2 ,{\bm k}_3}$ and 
$\theta_{{\bm k}_3 ,{\bm k}_1}$ are not necessarily close to $\pi /2$.
In other words, unless we take the strict squeezed limit ${\bm k}_1 \to -{\bm k}_2$, 
we can consider any angle between ${\bm k}_3$ and ${\bm k}_1, {\bm k}_2$. 
Averaging over $f_{\rm NL}$ in all the directions along the same 
line as Eq.~(\ref{sinave}), the local estimator for the squeezed shape
($k_3\ll k_1\simeq k_2$, $\theta_{{\bm k}_1, {\bm k}_3}\to \pi -\theta_{{\bm k}_2, {\bm k}_3}$, and $\theta_{{\bm k}_2, {\bm x}} \to \pi - \theta_{{\bm k}_1, {\bm x}}$) reads
\begin{equation}
f_{\rm NL}^{\rm local, average} \simeq 27\left(\frac{-g_*}{0.1}\right)\left(\frac{N_k}{60}\right)
\frac{[1- \cos^2 \theta_{{\bm k}_1, {\bm x}} -\cos^2\theta_{{\bm k}_3,{\bm x}}
+ \cos\theta_{{\bm k}_1, {\bm x}}\cos\theta_{{\bm k}_3, {\bm x}}\cos\theta_{{\bm k}_1, {\bm k}_3}]}{4/9} \,,
\label{fnl_vec_local}
\end{equation}
where we used the fact that the average value of the function 
in the last square bracket integrated over all the angles is $4/9$.
In this analysis the local non-linear estimator does not vanish and 
it can be as large as the order of $10$ for $g_* \sim -0.1$.

The Planck group placed the bound on the local non-linear estimator 
to be $f_{\rm NL}^{\rm local} = 2.7 \pm 5.8$ (68\,\%\,CL) \cite{Adenon}, 
but we cannot literally use this bound to constrain the anisotropic 
inflationary models. As we have seen above, the local non-Gaussianities
depend on what kind of squeezed shapes to be taken in the data analysis. 
It is interesting that the strict or incomplete squeezed shapes do matter 
to estimate the level of non-Gaussianities correctly.
The detailed data analysis based on different squeezed shapes is
beyond the scope of our paper.

Using the result (\ref{fnl_vec2}), we can calculate the non-linear estimator for 
the shapes other than the squeezed one.
In the equilateral ($\beta\rightarrow\pi/3\,, r_3\rightarrow1$)
and the enfolded ($\beta\rightarrow\pi \,, r_3\rightarrow2$) limits, we have
\begin{eqnarray}
f_{\rm NL}^{\rm equil} &\simeq& 7.5 \left( \frac{-g_*}{0.1} \right)\left( \frac{N_k}{60} \right) \,,\\
f_{\rm NL}^{\rm enfolded} &\simeq& 60 \left( \frac{-g_*}{0.1} \right)\left( \frac{N_k}{60} \right) \sin^2\gamma \,.
\end{eqnarray}
Unlike the squeezed case, these estimators do not depend on 
how we take the limits of equilateral and enfolded shapes.
The equilateral non-linear estimator is independent of the angle $\gamma$ 
and it is typically of the order of $1$ for $|g_*| \lesssim 0.1$.
Meanwhile, $f_{\rm NL}^{\rm enfolded}$ depends on $\gamma$ and 
it has a maximum at $\gamma=\pi/2$. 
The maximum value of $f_{\rm NL}^{\rm enfolded}$ can be as large as 
$60$ for $|g_*| \sim 0.1$, which is an interesting signature of the vector model.

%%%%%%%%%%%%%%%%%%%%%%%%%%%%%%%%%%%%
\subsection{$f(\phi)^2H_{\mu \nu \lambda}H^{\mu \nu \lambda}$ model}
%%%%%%%%%%%%%%%%%%%%%%%%%%%%%%%%%%%%

In the two-form model the Hamiltonian following from the third-order 
interacting Lagrangian $L_{\rm int}^{(3)}\simeq a^4 \delta E_{ij} \delta E_{ij} \zeta /2$
is given by 
\begin{equation} 
H_{\zeta2}=-\frac{1}{2H^4\tau^4}\int \, \frac{d^3k d^3p}{(2\pi)^{3/2}} \, 
\delta{\cal E}_{ij}({\bm k},\tau) \delta {\cal E}_{ij}({\bm p},\tau) 
\hat{\zeta}^{(0)}(-{\bm k}-{\bm p},\tau) \,.
\label{Hzeta2_two}
\end{equation}
The three-point correlation function of $\zeta$ can be computed by using the 
interacting Hamiltonians (\ref{Hzeta2}) and (\ref{Hzeta2_two}) in the way similar
to the derivation of Eq.~(\ref{bispe}). 
In Ref.~\cite{Ohashi} this was already derived as 
\begin{equation}
\delta \langle 0|\hat{\zeta}({\bm k}_1) \hat{\zeta}({\bm k}_2)\hat{\zeta}({\bm k}_3) |0 \rangle 
= 36\sqrt{2} \pi^{5/2}\frac{E_{yz}^2}{\epsilon V} N_{k_1} N_{k_2} N_{k_3} 
({\cal P}_{\zeta}^{(0)})^2
\delta^{(3)} ({\bm k}_1 + {\bm k}_2 + {\bm k}_3) 
\left[\frac{\cos \theta_{{\bm k}_1,{\bm k}_2} \cos\theta_{{\bm k}_1,{\bm x}} 
\cos\theta_{{\bm k}_2, {\bm x}}}{k_1^3k_2^3} + \text{2 perm.}\right].
\label{bispe_two}
\end{equation}
{}From the definition (\ref{def_fnl}) of $f_{\rm NL}$, it follows that 
\begin{eqnarray}
f_{\rm NL}& \simeq &30\left(\frac{g_*}{0.1}\right)\left(\frac{N_k}{60}\right)\frac{1}{k_1^3+k_2^3+k_3^3} 
[k_3^3 \cos \theta_{{\bm k}_1,{\bm k}_2} 
\cos\theta_{{\bm k}_1,{\bm x}} \cos\theta_{{\bm k}_2, {\bm x}}
+ k_1^3\cos \theta_{{\bm k}_2,{\bm k}_3} 
\cos\theta_{{\bm k}_2,{\bm x}} \cos\theta_{{\bm k}_3, {\bm x}} \nonumber \\
& &~~~~~~~~~~~~~~~~~~~~~~~~~~~~~~~~~~~~~~~
+k_2^3\cos \theta_{{\bm k}_3,{\bm k}_1} 
\cos\theta_{{\bm k}_3,{\bm x}} \cos\theta_{{\bm k}_1, {\bm x}}] \,,
\label{fnl_two}
\end{eqnarray}
where we used the anisotropic parameter $g_*$ given in 
Eq.~(\ref{g_two}) with $r_B=E_{yz}^2/(2\epsilon V)$.

For the squeezed shape, if we take the average over angles as 
the derivation of Eq.~(\ref{fnl_vec_local}), the local non-linear 
parameter reads
\begin{equation}
f_{\rm NL}^{\rm local,average} \simeq 
3.3\left(\frac{g_*}{0.1}\right)\left(\frac{N_k}{60}\right)
\frac{[\cos\theta_{{\bm k}_1, {\bm x}}\cos\theta_{{\bm k}_3, {\bm x}}
\cos\theta_{{\bm k}_1, {\bm k}_3}]}{1/9} \,,
\label{fnl_two_local}
\end{equation}
where the value $1/9$ is the total spatial average of the function 
in the last square bracket.
Therefore, $f_{\rm NL}^{\rm local,average}$ dose not vanish in this analysis.
We note that $f_{\rm NL}^{\rm local,average}$ in this case 
is smaller than that in the vector model by one order of magnitude, see 
Eq.~(\ref{fnl_vec_local}). Thus the local non-Gaussianities averaged over 
all the directions can allow us to discriminate between the vector and the two-form models.
On the other hand, in the strict squeezed limit characterized by 
$k_3 \to 0$, $ \theta_{{\bm k}_2,{\bm k}_3} \to \pi/2$, and 
$\theta_{{\bm k}_3,{\bm k}_1} \to \pi/2$, the non-linear estimator 
(\ref{fnl_two}) vanishes, $f_{\rm NL}^{\rm local}=0$ \cite{Ohashi}.

The non-linear parameters in the equilateral and the enfolded limits 
were already evaluated in Ref.~\cite{Ohashi} by considering a configuration 
with $k_1=k_2$ and $\beta=\pi-\theta_{{\bm k}_1, {\bm k}_2}$, 
$\gamma=\theta_{{\bm k}_1, {\bm x}}$.
They are given, respectively, by 
\begin{eqnarray}
f_{\rm NL}^{\rm equil} &\simeq& 3.7 \left( \frac{g_*}{0.1} \right)\left( \frac{N_k}{60} \right) \,,\\
f_{\rm NL}^{\rm enfolded} &\simeq& 30 \left( \frac{g_*}{0.1} \right)\left( \frac{N_k}{60} \right) \cos^2\gamma \,.
\end{eqnarray}
The equilateral non-linear estimator is independent of $\gamma$, but the 
enfolded one depends on $\gamma$. 
Unlike the vector model, $f_{\rm NL}^{\rm enfolded}$
has a maximum at $\gamma=0$ or $\pi$.
Both $f_{\rm NL}^{\rm equil}$ and $f_{\rm NL}^{\rm enfolded}$ are about 
half times smaller than those in the vector model.
In general, for the same values of $|g_*|$, 
the two-form model is easier to satisfy observational bounds
of non-Gaussianities relative to the vector model.

%===============================================================%
%************************ SECTION VI ****************************%
%===============================================================%

%%%%%%%%%%%
\section{Conclusions}
\label{consec}
%%%%%%%%%%%

We have studied differences between two anisotropic inflationary scenarios 
paying particular attention to their observational signatures. 
In the presence of a vector or a two-form field coupled to the inflaton, 
there exist background solutions along which the anisotropic hairs 
survive during inflation.
In the vector model the anisotropic shear $\Sigma$ divided by 
the Hubble rate $H$ is given by Eq.~(\ref{SigH1}) for the coupling (\ref{fphive}), 
while in the two-form model it is given by Eq.~(\ref{sigtwo2}) 
for the coupling (\ref{coutwo}). 
The opposite signs of $\Sigma/H$ between the two models 
reflect the fact that the types of anisotropies are different 
(either oblate or prolate).

In the presence of anisotropic interactions, 
we have computed the power spectra of scalar/tensor 
perturbations and the resulting spectral indices convenient to 
confront with the CMB observations. 
The different types of anisotropies affect the sign of $g_*$ 
appearing in the scalar power spectrum defined by Eq.~(\ref{anispe}), i.e., 
$g_*<0$ in the vector model and $g_*>0$ in the two-form model. 
The anisotropic contributions to the isotropic tensor power spectrum 
are suppressed relative to those to the isotropic 
scalar one. In particular, in the two-form model, 
we showed that anisotropic interactions do not give 
rise to any corrections to the isotropic tensor power spectrum.

We have computed the two-point cross correlations between 
curvature perturbations and gravitational waves.
While the cross correlation remains in the vector model, we find that 
it vanishes in the two-form model. The latter property may reflect the 
fact that, unlike the vector field, the two-form field can be mapped 
to the form of a scalar field.
The different signatures of the two anisotropic models will be 
useful to discriminate between those models in future observations 
of the TB power spectrum.

In the light of the recent Planck data, we have placed observational 
constraints on several different inflaton potentials from the information 
of the scalar spectral index $n_s$ and the tensor-to-scalar ratio $r$. 
In both the vector and two-form models anisotropic interactions 
lead to the enhancement of the scalar power spectrum on larger 
scales, by which both $n_s$ and $r$ decrease for any inflaton 
potentials. In the vector model, we found that the potentials 
$V(\phi)=\lambda \phi$ and $V(\phi)=(3\lambda/2)\phi^{2/3}$ show 
the better compatibility with the data for larger $|g_*|$ and 
that the potential of natural inflation is outside the 
68\,\%\,CL region constrained by the Planck+WP+BAO+high-$\ell$ 
data for $|g_*| \gtrsim 0.5$. 
In the two-form model, the level of the decreases of $n_s$ 
and $r$ is less significant relative to the vector model 
for the same values of $|g_*|$. 
The potential (\ref{pot_S}), which originates from the 
Starobinsky's $f(R)$ model, is well 
inside the 68\,\%\,CL region even in the presence of 
anisotropic interactions with $|g_*|<0.5$. 
 
We also compared the three-point correlation functions of curvature 
perturbations between the two anisotropic inflationary scenarios.
If the strict squeezed limit characterized by $k_3 \to 0$, 
$\theta_{{\bm k}_1 ,{\bm k}_3} \to \pi/2$, and $\theta_{{\bm k}_2 ,{\bm k}_3} \to \pi/2$ 
is taken, the local non-linear estimators $f_{\rm NL}^{\rm local}$ 
vanish in both the vector and the two-form models. 
However, for the squeezed shape where the angles 
$\theta_{{\bm k}_1 ,{\bm k}_3}$ and $\theta_{{\bm k}_2 ,{\bm k}_3}$ 
are not necessarily close to $\pi/2$, the non-linear estimator averaged
over all the directions is given by Eq.~(\ref{fnl_vec_local}) in the vector model and 
by Eq.~(\ref{fnl_two_local}) in the two-form model. 
The former is larger than the latter by one order of magnitude 
for the same order of the anisotropic parameter $g_*$.
We also evaluated the non-linear estimators in the equilateral and 
enfolded limits and found that $f_{\rm NL}$ in the vector model is about twice 
larger than that in the two-form model for the same values of $|g_*|$. 
In general, the two-form model is easier to satisfy observational 
bounds of non-Gaussianities relative to the vector model.

We have thus shown that the two anisotropic inflationary scenarios can be 
distinguished from each other by evaluating several CMB observables.
In particular, the precise measurements of $g_*$ as well as 
the TB correlation will clarify which anisotropic model is 
favored over the other.

\acknowledgements

This work is supported by the Grant-in-Aid for Scientific Research 
Fund of the Ministry of Education, Science and Culture of Japan 
(Nos.~23$\cdot$6781, 25400251, and 24540286), the Grant-in-Aid 
for Scientific Research on Innovative Area (No.~21111006).


\begin{thebibliography}{99}

\bibitem{WMAP1} 
D.~N.~Spergel {\it et al.}  [WMAP Collaboration],
%``First year Wilkinson Microwave Anisotropy Probe (WMAP) observations: 
%Determination of cosmological parameters,''
Astrophys.\ J.\ Suppl.\  {\bf 148}, 175 (2003)
[astro-ph/0302209].

\bibitem{WMAP9} 
G.~Hinshaw {\it et al.}  [WMAP Collaboration],
%``Nine-Year Wilkinson Microwave Anisotropy Probe (WMAP) Observations: 
%Cosmological Parameter Results,''
arXiv:1212.5226 [astro-ph.CO].

\bibitem{Rei} 
C.~L.~Reichardt  {\it et al.},
%``A measurement of secondary cosmic microwave background 
%anisotropies with two years of South Pole Telescope observations,''
Astrophys.\ J.\  {\bf 755}, 70 (2012)
[arXiv:1111.0932 [astro-ph.CO]].

\bibitem{Atacama} 
J.~L.~Sievers {\it et al.},
%``The Atacama Cosmology Telescope: Cosmological parameters from three seasons of data,''
arXiv:1301.0824 [astro-ph.CO].

\bibitem{LSS1} 
M.~Tegmark {\it et al.}  [SDSS Collaboration],
%``Cosmological parameters from SDSS and WMAP,''
Phys.\ Rev.\ D {\bf 69}, 103501 (2004)
[astro-ph/0310723].

\bibitem{LSS2} 
C.~P.~Ahn {\it et al.}  [SDSS Collaboration],
%``The Ninth Data Release of the Sloan Digital Sky Survey: First Spectroscopic 
%Data from the SDSS-III Baryon Oscillation Spectroscopic Survey,''
 Astrophys.\ J.\ Suppl.\  {\bf 203}, 21 (2012)
[arXiv:1207.7137 [astro-ph.IM]].

\bibitem{Adecosmo} 
P.~A.~R.~Ade {\it et al.}  [Planck Collaboration],
%``Planck 2013 results. XVI. Cosmological parameters,''
arXiv:1303.5076 [astro-ph.CO].

\bibitem{Adeinf} 
P.~A.~R.~Ade {\it et al.}  [Planck Collaboration],
%``Planck 2013 results. XXII. Constraints on inflation,''
arXiv:1303.5082 [astro-ph.CO].

\bibitem{Adenon}
P.~A.~R.~Ade {\it et al.}  [ Planck Collaboration],
%``Planck 2013 Results. XXIV. Constraints on primordial non-Gaussianity,''
arXiv:1303.5084 [astro-ph.CO].

\bibitem{oldper} 
V.~F.~Mukhanov and G.~V.~Chibisov, 
%``Quantum Fluctuation And 'Nonsingular' Universe.''
JETP Lett.\ \textbf{33}, 532 (1981); \\
A.~H.~Guth and S.~Y.~Pi,
%``Fluctuations In The New Inflationary Universe,''
Phys.\ Rev.\ Lett.\ \textbf{49} (1982) 1110; \\
S.~W.~Hawking, Phys.\ Lett.\ B \textbf{115}, 295 (1982); \\
A.~A.~Starobinsky, 
%``Dynamics Of Phase Transition In The New Inflationary
%Universe Scenario And Generation Of Perturbations,''
Phys.\ Lett.\ B \textbf{117} (1982) 175; \\
J.~M.~Bardeen, P.~J.~Steinhardt and M.~S.~Turner, 
Phys.\ Rev.\ D \textbf{28}, 679 (1983).

\bibitem{aniobser1} 
N.~E.~Groeneboom and H.~K.~Eriksen,
%``Bayesian analysis of sparse anisotropic universe 
%models and application to the 5-yr WMAP data,''
Astrophys.\ J.\  {\bf 690}, 1807 (2009)
[arXiv:0807.2242 [astro-ph]].

\bibitem{aniobser2} 
D.~Hanson and A.~Lewis,
%``Estimators for CMB Statistical Anisotropy,''
Phys.\ Rev.\ D {\bf 80}, 063004 (2009)
[arXiv:0908.0963 [astro-ph.CO]].

\bibitem{aniobser3} 
N.~E.~Groeneboom, L.~Ackerman, I.~K.~Wehus and H.~K.~Eriksen,
%``Bayesian analysis of an anisotropic universe model: systematics and polarization,''
Astrophys.\ J.\  {\bf 722}, 452 (2010)
[arXiv:0911.0150 [astro-ph.CO]].

\bibitem{Carroll}
L.~Ackerman, S.~M.~Carroll and M.~B.~Wise,
%``Imprints of a Primordial Preferred Direction on the Microwave Background,''
Phys.\ Rev.\ D {\bf 75}, 083502 (2007)
[Erratum-ibid.\ D {\bf 80}, 069901 (2009)]
[astro-ph/0701357].

\bibitem{Hanson}
D.~Hanson, A.~Lewis and A.~Challinor,
%``Asymmetric Beams and CMB Statistical Anisotropy,''
Phys.\ Rev.\ D {\bf 81}, 103003 (2010)
[arXiv:1003.0198 [astro-ph.CO]].

\bibitem{Kim}
J.~Kim and E.~Komatsu,
arXiv:1310.1605 [astro-ph.CO].

\bibitem{review1} 
J.~Soda,
%``Statistical Anisotropy from Anisotropic Inflation,''
Class.\ Quant.\ Grav.\  {\bf 29}, 083001 (2012)
[arXiv:1201.6434 [hep-th]].

\bibitem{review2} 
A.~Maleknejad, M.~M.~Sheikh-Jabbari and J.~Soda,
%``Gauge Fields and Inflation,''
Phys.\ Rept.\  {\bf 528}, 161 (2013)
[arXiv:1212.2921 [hep-th]].
  
\bibitem{Watanabe}
M.~a.~Watanabe, S.~Kanno and J.~Soda,
%``Inflationary Universe with Anisotropic Hair,''
Phys.\ Rev.\ Lett.\  {\bf 102}, 191302 (2009)
[arXiv:0902.2833 [hep-th]].

\bibitem{early}
S.~Yokoyama and J.~Soda,
%``Primordial statistical anisotropy generated at the end of inflation,''
JCAP {\bf 0808}, 005 (2008)
[arXiv:0805.4265 [astro-ph]];\\
K.~Dimopoulos, M.~Karciauskas, D.~H.~Lyth and Y.~Rodriguez,
%``Statistical anisotropy of the curvature perturbation from vector field
%perturbations,''
JCAP {\bf 0905}, 013 (2009)
[arXiv:0809.1055 [astro-ph]];\\
M.~Karciauskas, K.~Dimopoulos and D.~H.~Lyth,
%``Anisotropic non-Gaussianity from vector field perturbations,''
Phys.\ Rev.\  D {\bf 80} (2009) 023509
[arXiv:0812.0264 [astro-ph]];\\
C.~A.~Valenzuela-Toledo, Y.~Rodriguez and D.~H.~Lyth,
%``Non-gaussianity at tree- and one-loop levels from vector field
%perturbations,''
Phys.\ Rev.\  D {\bf 80}, 103519 (2009)
[arXiv:0909.4064 [astro-ph.CO]];\\
S.~Kanno, J.~Soda and M.~a.~Watanabe,
%``Cosmological Magnetic Fields from Inflation and Backreaction,''
JCAP {\bf 0912}, 009 (2009)
[arXiv:0908.3509 [astro-ph.CO]];\\
C.~A.~Valenzuela-Toledo and Y.~Rodriguez,
%``Non-gaussianity from the trispectrum and vector field perturbations,''
Phys.\ Lett.\  B {\bf 685}, 120 (2010)
[arXiv:0910.4208 [astro-ph.CO]];\\
N.~Bartolo, E.~Dimastrogiovanni, S.~Matarrese and A.~Riotto,
%``Anisotropic bispectrum of curvature perturbations from primordial
%non-Abelian vector fields,''
JCAP {\bf 0910}, 015 (2009)
[arXiv:0906.4944 [astro-ph.CO]];\\
N.~Bartolo, E.~Dimastrogiovanni, S.~Matarrese and A.~Riotto,
%``Anisotropic Trispectrum of Curvature Perturbations Induced by Primordial
%Non-Abelian Vector Fields,''
JCAP {\bf 0911}, 028 (2009)
[arXiv:0909.5621 [astro-ph.CO]];\\
E.~Dimastrogiovanni, N.~Bartolo, S.~Matarrese and A.~Riotto,
%``Non-Gaussianity and statistical anisotropy from vector field populated
%inflationary models,''
Adv.\ Astron.\  {\bf 2010}, 752670 (2010)
[arXiv:1001.4049 [astro-ph.CO]];\\
J.~M.~Wagstaff and K.~Dimopoulos,
%``Particle Production of Vector Fields: Scale Invariance is Attractive,''
Phys.\ Rev.\  D {\bf 83}, 023523 (2011)
[arXiv:1011.2517 [hep-ph]].

\bibitem{powerlaw}
S.~Kanno, J.~Soda and M.~a.~Watanabe,
%``Anisotropic Power-law Inflation,''
JCAP {\bf 1012}, 024 (2010)
[arXiv:1010.5307 [hep-th]].

\bibitem{anipapers}
P.~V.~Moniz and J.~Ward, 
%``Gauge field back-reaction in Born Infeld cosmologies,''
Class.\ Quant.\ Grav.\  {\bf 27}, 235009 (2010)
[arXiv:1007.3299 [gr-qc]];\\
K.~Murata and J.~Soda,
%``Anisotropic Inflation with Non-Abelian Gauge Kinetic Function,''
JCAP\ {\bf 1106}, 037  (2011)
[arXiv:1103.6164 [hep-th]];\\
T.~Q.~Do, W.~F.~Kao and I.~-C.~Lin,
%``Anisotropic power-law inflation for a two scalar fields model,''
Phys.\ Rev.\ D\ {\bf 83}, 123002  (2011);\\
T.~Q.~Do and W.~F.~Kao,
%``Anisotropic power-law inflation for the Dirac-Born-Infeld theory,''
Phys.\ Rev.\  D {\bf 84}, 123009 (2011);\\
R.~Emami, H.~Firouzjahi, S.~M.~Sadegh Movahed and M.~Zarei,
%``Anisotropic Inflation from Charged Scalar Fields,''
JCAP {\bf 1102}, 005 (2011)
[arXiv:1010.5495 [astro-ph.CO]];\\
M.~Karciauskas,
%``The Primordial Curvature Perturbation from Vector Fields 
%of General non-Abelian Groups,''
JCAP {\bf 1201}, 014 (2012)
[arXiv:1104.3629 [astro-ph.CO]];\\
M.~Shiraishi and S.~Yokoyama,
%``Violation of the Rotational Invariance in the CMB Bispectrum,''
Prog.\ Theor.\ Phys.\  {\bf 126}, 923 (2011)
[arXiv:1107.0682 [astro-ph.CO]];\\
S.~Bhowmick and S.~Mukherji,
%``Anisotropic Power Law Inflation from Rolling Tachyons,''
Mod.\ Phys.\ Lett.\ A {\bf 27}, 1250009 (2012)
[arXiv:1105.4455 [hep-th]];\\
R.~Emami and H.~Firouzjahi,
%``Issues on Generating Primordial Anisotropies at the End of Inflation,''
JCAP {\bf 1201}, 022 (2012)
[arXiv:1111.1919 [astro-ph.CO]];\\
K.~Yamamoto, M.~a.~Watanabe and J.~Soda,
%``Inflation with Multi-Vector-Hair: The Fate of Anisotropy,''
Class.\ Quant.\ Grav.\  {\bf 29}, 145008 (2012)
[arXiv:1201.5309 [hep-th]];\\
K.~Dimopoulos,
%``Statistical Anisotropy and the Vector Curvaton Paradigm,''
Int.\ J.\ Mod.\ Phys.\ D {\bf 21}, 1250023 (2012)
[Erratum-ibid.\ D {\bf 21}, 1292003 (2012)]
[arXiv:1107.2779 [hep-ph]];\\
A.~A.~Abolhasani, R.~Emami, J.~T.~Firouzjaee and H.~Firouzjahi,
%``\delta N Formalism in Anisotropic Inflation and Large Anisotropic 
%Bispectrum and Trispectrum,''
arXiv:1302.6986 [astro-ph.CO];\\
D.~H.~Lyth and M.~Karciauskas,
%``The statistically anisotropic curvature perturbation 
%generated by f(\phi)^2 F^2,''
arXiv:1302.7304 [astro-ph.CO];\\
S.~Baghram, M.~H.~Namjoo and H.~Firouzjahi,
%``Large Scale Anisotropic Bias from Primordial non-Gaussianity,''
arXiv:1303.4368 [astro-ph.CO];\\
N.~Bartolo, S.~Matarrese, M.~Peloso and A.~Ricciardone,
%``Anisotropy in solid inflation,''
arXiv:1306.4160 [astro-ph.CO].
%%CITATION = ARXIV:1306.4160;%%

\bibitem{Gum} 
A.~E.~Gumrukcuoglu, B.~Himmetoglu and M.~Peloso,
%``Scalar-Scalar, Scalar-Tensor, and Tensor-Tensor Correlators
% from Anisotropic Inflation,''
Phys.\ Rev.\ D {\bf 81}, 063528 (2010)
[arXiv:1001.4088 [astro-ph.CO]];\\
T.~R.~Dulaney and M.~I.~Gresham,
%``Primordial Power Spectra from Anisotropic Inflation,''
Phys.\ Rev.\ D {\bf 81}, 103532 (2010)
[arXiv:1001.2301 [astro-ph.CO]].

\bibitem{Watanabe:2010fh}
M.~a.~Watanabe, S.~Kanno and J.~Soda,
%``The Nature of Primordial Fluctuations from Anisotropic Inflation,''
Prog.\ Theor.\ Phys.\  {\bf 123}, 1041 (2010)
[arXiv:1003.0056 [astro-ph.CO]].  

\bibitem{Bartolo} 
N.~Bartolo, S.~Matarrese, M.~Peloso and A.~Ricciardone,
%``The anisotropic power spectrum and bispectrum in the f(phi) F^2 mechanism,''
Phys.\ Rev.\ D {\bf 87}, 023504 (2013)
[arXiv:1210.3257 [astro-ph.CO]].

\bibitem{Shiraishi} 
M.~Shiraishi, E.~Komatsu, M.~Peloso and N.~Barnaby,
%``Signatures of anisotropic sources in the squeezed-limit bispectrum 
%of the cosmic microwave background,''
JCAP {\bf 1305}, 002 (2013)
[arXiv:1302.3056 [astro-ph.CO]].

\bibitem{Barnaby} 
N.~Barnaby and M.~Peloso, 
%``Large Nongaussianity in Axion Inflation,''
Phys.\ Rev.\ Lett.\  {\bf 106}, 181301 (2011)
[arXiv:1011.1500 [hep-ph]];\\
N.~Barnaby, R.~Namba and M.~Peloso,
%``Phenomenology of a Pseudo-Scalar Inflaton: Naturally Large Nongaussianity,''
JCAP {\bf 1104}, 009 (2011)
[arXiv:1102.4333 [astro-ph.CO]];\\
N.~Barnaby, E.~Pajer and M.~Peloso, 
%``Gauge Field Production in Axion Inflation: Consequences for 
%Monodromy, non-Gaussianity in the CMB, and Gravitational 
%Waves at Interferometers,''
Phys.\ Rev.\ D {\bf 85}, 023525 (2012)
[arXiv:1110.3327 [astro-ph.CO]];\\
N.~Barnaby, R.~Namba and M.~Peloso,
%``Observable non-gaussianity from gauge field production 
%in slow roll inflation, and a challenging connection with magnetogenesis,''
Phys.\ Rev.\ D {\bf 85}, 123523 (2012)
[arXiv:1202.1469 [astro-ph.CO]].

\bibitem{Ohashi} 
J.~Ohashi, J.~Soda and S.~Tsujikawa,
%``Anisotropic Non-Gaussianity from a Two-Form Field,''
Phys.\ Rev.\ D {\bf 87}, 083520 (2013)
[arXiv:1303.7340 [astro-ph.CO]].

\bibitem{Maldacena:2002vr}
J.~M.~Maldacena,
%``Non-Gaussian features of primordial fluctuations in single field
%inflationary models,''
JHEP {\bf 0305}, 013 (2003).

\bibitem{Kuro} 
S.~Tsujikawa, J.~Ohashi, S.~Kuroyanagi and A.~De Felice,
%``Planck constraints on single-field inflation,''
Phys.\  Rev.\ D {\bf 88}, 023529 (2013)
[arXiv:1305.3044 [astro-ph.CO]].

\bibitem{Watanabe2}
M.~a.~Watanabe, S.~Kanno and J.~Soda,
%``Imprints of anisotropic inflation on the cosmic microwave background,''
Mon.\ Not.\ Roy.\ Astron.\ Soc.\  {\bf 412}, L83 (2011)
[arXiv:1011.3604 [astro-ph.CO]].

\bibitem{Hervik}
S.~Hervik, D.~F.~Mota and M.~Thorsrud,
%``Inflation with stable anisotropic hair: Is it cosmologically viable?,''
JHEP {\bf 1111}, 146 (2011)
[arXiv:1109.3456 [gr-qc]].

\bibitem{kinf} 
C.~Armendariz-Picon, T.~Damour and V.~F.~Mukhanov,
%``k - inflation,''
Phys.\ Lett.\ B {\bf 458}, 209 (1999)
[hep-th/9904075].

\bibitem{anikinf} 
J.~Ohashi, J.~Soda and S.~Tsujikawa,
%``Anisotropic power-law k-inflation,''
arXiv:1310.3053 [hep-th].

\bibitem{Bardeen} 
J.~M.~Bardeen,
%``Gauge Invariant Cosmological Perturbations,''
Phys.\ Rev.\ D {\bf 22}, 1882 (1980).

\bibitem{Bardeen2} 
J.~M.~Bardeen, P.~J.~Steinhardt and M.~S.~Turner,
%``Spontaneous Creation of Almost Scale - Free Density Perturbations in an Inflationary Universe,''
Phys.\ Rev.\ D {\bf 28}, 679 (1983);\\
V.~N.~Lukash,
%``Production of phonons in an isotropic universe,''
Sov.\ Phys.\ JETP {\bf 52}, 807 (1980).

\bibitem{zetadef} 
V.~F.~Mukhanov, H.~A.~Feldman and R.~H.~Brandenberger,
%``Theory of cosmological perturbations. Part 1. Classical 
%perturbations. Part 2. Quantum theory of perturbations. Part 3. Extensions,''
Phys.\ Rept.\  {\bf 215}, 203 (1992);\\
B.~A.~Bassett, S.~Tsujikawa and D.~Wands,
%``Inflation dynamics and reheating,''
Rev.\ Mod.\ Phys.\  {\bf 78}, 537 (2006)
[astro-ph/0507632].

\bibitem{Horn} 
A.~De Felice and S.~Tsujikawa,
%``Shapes of primordial non-Gaussianities in the Horndeski's 
%most general scalar-tensor theories,''
JCAP {\bf 1303}, 030 (2013)
[arXiv:1301.5721 [hep-th]].

\bibitem{Soda:2011am} 
J.~Soda, H.~Kodama and M.~Nozawa,
%``Parity Violation in Graviton Non-gaussianity,''
JHEP {\bf 1108}, 067 (2011)
[arXiv:1106.3228 [hep-th]].
%%CITATION = ARXIV:1106.3228;%%
%16 citations counted in INSPIRE as of 31 Jul 2013

\bibitem{cosmomc}
http://cosmologist.info/cosmomc/ 

\bibitem{Lewis}
A.~Lewis,
%``Efficient sampling of fast and slow cosmological parameters,''
Phys.\  Rev.\  D87, {\bf 103529} (2013)
[arXiv:1304.4473 [astro-ph.CO]].

\bibitem{BAO1} 
F.~Beutler {\it et al.},
%``The 6dF Galaxy Survey: Baryon Acoustic Oscillations and 
%the Local Hubble Constant,''
Mon.\ Not.\ Roy.\ Astron.\ Soc.\  {\bf 416}, 3017 (2011)
[arXiv:1106.3366 [astro-ph.CO]].

\bibitem{BAO2} 
N.~Padmanabhan  {\it et al.},
%``A 2% Distance to z=0.35 by Reconstructing Baryon Acoustic 
%Oscillations - I : Methods and Application to the Sloan Digital Sky Survey,''
arXiv:1202.0090 [astro-ph.CO].

\bibitem{BAO3} 
L.~Anderson {\it et al.},
%``The clustering of galaxies in the SDSS-III Baryon Oscillation 
%Spectroscopic Survey: Baryon Acoustic Oscillations in the Data 
%Release 9 Spectroscopic Galaxy Sample,''
Mon.\ Not.\ Roy.\ Astron.\ Soc.\  {\bf 427}, no. 4, 3435 (2013)
[arXiv:1203.6594 [astro-ph.CO]].

\bibitem{Das} 
S.~Das {\it et al.},
%``The Atacama Cosmology Telescope: Temperature and Gravitational 
%Lensing Power Spectrum Measurements from Three Seasons of Data,''
arXiv:1301.1037 [astro-ph.CO].

\bibitem{chaotic} 
A.~D.~Linde,
%``Chaotic Inflation,''
Phys.\ Lett.\ B {\bf 129}, 177 (1983).

\bibitem{mono1} 
L.~McAllister, E.~Silverstein and A.~Westphal,
%``Gravity Waves and Linear Inflation from Axion Monodromy,''
Phys.\ Rev.\ D {\bf 82}, 046003 (2010)
[arXiv:0808.0706 [hep-th]].

\bibitem{mono2} 
E.~Silverstein and A.~Westphal,
%``Monodromy in the CMB: Gravity Waves and String Inflation,''
Phys.\ Rev.\ D {\bf 78}, 106003 (2008)
[arXiv:0803.3085 [hep-th]].

\bibitem{Freese} 
K.~Freese, J.~A.~Frieman and A.~V.~Olinto,
%``Natural inflation with pseudo - Nambu-Goldstone bosons,''
Phys.\ Rev.\ Lett.\  {\bf 65}, 3233 (1990);\\
F.~C.~Adams, J.~R.~Bond, K.~Freese, J.~A.~Frieman and A.~V.~Olinto,
%``Natural inflation: Particle physics models, power law spectra 
%for large scale structure, and constraints from COBE,''
Phys.\ Rev.\ D {\bf 47}, 426 (1993).

\bibitem{Starobinsky} 
A.~A.~Starobinsky,
%``A New Type of Isotropic Cosmological Models Without Singularity,''
Phys.\ Lett.\ B {\bf 91}, 99 (1980).

\bibitem{Maeda} 
K.~-i.~Maeda,
%``Towards the Einstein-Hilbert Action via Conformal Transformation,''
Phys.\ Rev.\ D {\bf 39}, 3159 (1989).

\bibitem{Ketov} 
S.~V.~Ketov and A.~A.~Starobinsky,
%``Embedding (R+R^2)-Inflation into Supergravity,''
Phys.\ Rev.\ D {\bf 83}, 063512 (2011);\\
S.~V.~Ketov and S.~Tsujikawa,
%``Consistency of inflation and preheating in F(R) supergravity,''
Phys.\ Rev.\ D {\bf 86}, 023529 (2012)
[arXiv:1205.2918 [hep-th]];\\
J.~Ellis, D.~V.~Nanopoulos and K.~A.~Olive,
%``A No-Scale Supergravity Realization of the Starobinsky Model,''
arXiv:1305.1247 [hep-th];\\
R.~Kallosh and A.~Linde,
%``Superconformal generalizations of the Starobinsky model,''
JCAP {\bf 1306}, 028 (2013)
[arXiv:1306.3214 [hep-th]];\\
F.~Farakos, A.~Kehagias and A.~Riotto,
%``On the Starobinsky Model of Inflation from Supergravity,''
arXiv:1307.1137 [hep-th];\\
W.~Buchmuller, V.~Domcke and K.~Kamada,
%``The Starobinsky Model from Superconformal D-Term Inflation,''
arXiv:1306.3471 [hep-th];\\
F.~Briscese, L.~Modesto and S.~Tsujikawa,
%``Super-renormalizable or finite completion of the Starobinsky theory,''
arXiv:1308.1413 [hep-th].

\bibitem{DeFelice} 
A.~De Felice and S.~Tsujikawa,
%``f(R) theories,''
Living Rev.\ Rel.\  {\bf 13}, 3 (2010)
[arXiv:1002.4928 [gr-qc]].

\bibitem{Staper} 
A.~A.~Starobinsky,
%``The Perturbation Spectrum Evolving from a Nonsingular Initially 
%De-Sitte r Cosmology and the Microwave Background Anisotropy,''
Sov.\ Astron.\ Lett.\  {\bf 9}, 302 (1983);\\
L.~A.~Kofman, V.~F.~Mukhanov and D.~Y.~Pogosian,
%``EVOLUTION OF INHOMOGENEITIES IN INFLATIONARY MODELS IN A THEORY OF
%GRAVITATION WITH HIGHER DERIVATIVES,''
Sov.\ Phys.\ JETP {\bf 66}, 433 (1987);\\
J.~c.~Hwang and H.~Noh,
%``$f(R)$ gravity theory and CMBR constraints,''
Phys.\ Lett.\  B {\bf 506}, 13 (2001).

\bibitem{hybrid} 
A.~D.~Linde,
%``Hybrid inflation,''
Phys.\ Rev.\ D {\bf 49}, 748 (1994)
[astro-ph/9307002].

\end{thebibliography}
\end{document}